\documentclass{article}
\usepackage{graphicx} 
\usepackage{authblk}
\usepackage[utf8]{inputenc}    
\usepackage[T1]{fontenc}       
\usepackage[english]{babel}    
\usepackage{lmodern} 
\usepackage{booktabs}
\usepackage{caption} 
\usepackage{float}
\usepackage{multirow}
\usepackage{hhline}
\usepackage{amssymb}
\usepackage{graphicx}
\usepackage{mathtools}
\usepackage{bm}
\usepackage{natbib}
\setcitestyle{authoryear,open={(},close={)}}
\usepackage{float}
\usepackage[hidelinks]{hyperref}

\title{Preplay Losing Contracts: Inducing Strong Nash Equilibrium in the $n$-player Prisoner's Dilemma}

\author{Ian Fligler}
\affil{Department of Mathematics \\ University of Buenos Aires \\ ianfligler@gmail.com}
\date{April 2026} 

\usepackage{indentfirst}
\begin{document}
\maketitle
\vspace{1\baselineskip}

\begin{abstract}
In strategic games such as the prisoner's dilemma, allowing players to
make binding offers of utility transfers before play has been shown to
alter incentives and potentially support cooperative outcomes. These
preplay exchange mechanisms reshape payoffs by transferring utility
while being contingent on actions; however, they typically require side
payments that can reduce individual benefits relative to joint
cooperation. In this paper, we extend the analysis to a finite
\(n\)-player prisoner's dilemma with ordered strategy sets, defined such
that any restriction of strategies by any subset of players still yields
a prisoner's dilemma. To achieve a robust cooperative outcome
that resists group deviations, we introduce a novel class of mechanisms: \textit{losing contracts}. Unlike transfer-based preplay mechanisms, losing
contracts require players to irrevocably reduce their own utility if
they defect, thereby aligning individual incentives with cooperation
without inter-player payments. With appropriately chosen loss amounts,
losing contracts induce joint cooperation as the unique strong Nash
equilibrium in the modified game and in every restricted game within
it, ensuring that cooperative incentives persist even under possible
external constraints on strategy sets. We show that our contracts can be constructively
defined, reducing the preplay stage to a simple and binary
decision for each player: whether to sign the contract or not. Furthermore, if the losing contract is only executed when all players
sign, signing is a strictly dominant strategy for all. Finally, we extend these
results to certain public goods games. 
\par\vspace{1em}
\noindent \textbf{Keywords}: Losing Contracts, $n$-player Prisoner's Dilemma, Strong Nash Equilibrium, Restricted strategy sets, Burning money \\
\noindent \textbf{JEL Classification}: C72, D71, D86
\end{abstract}

\newpage

\tableofcontents
\clearpage

\section{Introduction}

The prisoner's dilemma is characterized by the discrepancy between the
ga\-me's Nash equilibrium and Pareto-optimal profile. Andreoni and Varian (\citeyear{AndreoniVarian1999}) propose a solution to this inconsistency, involving the
introduction of a pre-game instance in which each player offers an
incentive to the other player that is contingent on their cooperation.
This mechanism can induce joint cooperation to become a Nash
equilibrium, although a player may be compelled to relinquish a
substantial portion of their utility with respect to the original game
in some instances. To circumvent this predicament, Qin (\citeyear{Qin2008}) suggests a
system in which payoffs are offered in the event of each player's
deviation. Consequently, the benefits of mutual cooperation remain
unaltered in relation to the original dilemma matrix. However, as in the
first model, it is challenging to extend this approach to
\emph{n}-player games, because this involves contemplating all possible
distribution combinations of each of the payoffs among the other
participants in the game.

This paper has two main purposes: first, to establish the mathematical
conditions of a prisoner's dilemma of \emph{n} players with finite
strategies each; second, to define a prior contract that not only
manages to optimize the cooperative equilibrium of the previous game but
also makes this equilibrium unique and strong in the Nash sense, even
when transferable utilities are considered. Consequently, the concept of
\textit{losing contracts} is introduced, aligning with the me\-chanism proposed by
Ben-Porath and Dekel (\citeyear{BenPorathDekel1992}). In these contracts, instead of offering
their resources to others as a reward, players consume a portion of
their resources. While these contracts result in the inefficient
strategy profile of game resources, it will become evident that they are the
only ones that consistently attain a robust cooperation equilibrium.

First, it is necessary to demonstrate that the prisoner's dilemma of
\emph{n} players with finite strategies acquires the tension of the
classical dilemma and that it does not exhibit coalitional incentives as
long as its cooperative equilibrium is Pareto-optimal. This renders our
subsequent analysis meaningful. Second, we demonstrate that the
definition of a losing contract can be established in such a manner that
joint cooperation constitutes a strong Nash equilibrium in every
constrained game. This result indicates that cooperating is the optimal
choice, even in scenarios where restrictions are imposed on one's or
other players' strategies. Moreover, this conceptualization of
losing contracts effectively reduces the pre-game instance to a binary
decision. It has also been determined that when the contract is executed
only if all players sign, signing emerges as a dominant strategy for all
involved parties. 

The rest of the work is organized as follows. Section 2 outlines some
preliminary definitions. Section 3 formally defines losing contracts and
discusses their advantages. Section 4 presents the
disadvantages of existing exchange contracts. Section 5 introduces the
\emph{n}-player prisoner's dilemma, justifies the motivation behind this
definition, and presents the main results. Finally, it extends the
findings to a weak version equivalent to certain public goods games. Section 6 concludes the paper.

\section{Preliminaries}

Consider a game in strategic form
\(G = \left\langle N,\left( S_{i} \right)_{i \in N},\left( u_{i} \right)_{i \in N} \right\rangle\) (we may write $G$ $=\left\langle N,\left( S_{i} \right)_{i \in N} \right\rangle$ if $(u_i)_{i\in N}$ is clear from context),
where \(N = \{ 1,\ldots,n\}\), \(S_{i}\) is the strategy set of player
\(i\), \(S = S_{1} \times \ldots \times S_{n}\) is the set of strategy profiles, and
\(u_{i}:S \rightarrow \mathbb{R}\) is the utility function of player \(i\). We denote by \(u_{i}\left( k_{1},\ldots,k_{n} \right)\) the value that represents the utility of player \(i\)
for the strategy profile
\(\left( s_{k_{1}},\ldots,s_{k_{n}} \right)\).

For any non-empty subset of players \(A \subseteq N\), \(S_{A}\) is the
set of member strategy profiles of \(A\).
\(s=\left( s_{A},s_{- A} \right) \in S\) is a strategy profile, where
\(s_{A} \in S_{A}\) represents the strategy profile of the group \(A\)
and \(s_{- A} \in S_{- A}\) of the remaining players.

A strategy profile \(s \in S\) is called Pareto-optimal under the transferable
utility assumption if it maximizes the total sum of all players'
utilities; that is, if
\(\sum_{i = 1}^{n}\mspace{2mu} u_{i}(\tilde{s}) \leq \sum_{i = 1}^{n}\mspace{2mu} u_{i}(s)\) \(\forall\tilde{s} \in S\).
This stronger Pareto-optimality condition is adopted due to the assumption that side payments are permitted.

\(\tilde{G}\) is an induced game from \(G\) if strategies are fixed
for some non-empty subset of players \(A \subseteq N\) and if
\(\tilde{G}\) contains two or more players whose strategies are not
fixed. \(G_{- A}\left( s_{A} \right)\) is the game induced by strategy profile
\(s_{A} \in S_{A}\) played by the members of \(N \setminus A\).

Under a transferable utility context, a strategy profile
\(s \in S\) is a strong Nash equilibrium if
\(\sum_{i \in A}^{}\mspace{2mu} u_{i}\left( {\tilde{s}}_{A},s_{- A} \right) \leq \sum_{i \in A}^{}\mspace{2mu} u_{i}(s)\)
for every non-empty coalition \(A \subseteq N\) and for every deviation
\({\tilde{s}}_{A} \in S_{A} \setminus \left\{ s_{A} \right\}\).
This equilibrium is strict if the above inequality is strict
\(\forall A \subseteq N,{\tilde{s}}_{A} \in S_{A} \setminus \left\{ s_{A} \right\}\).
Otherwise, the result is not considered to be strict. Unless otherwise stated, any strong Nash equilibrium will be assumed to be strict.

\section{Losing contracts}

\subsection{Illustrative example}

Consider the following version of the prisoner's dilemma. In the event
that both suspects confess, they will each serve a five-year prison
sentence; if neither confesses, they will each serve a one-year
sentence; and if one confesses and the other does not, the one who
confesses will be released while the one who does not confess will serve
the full ten-year sentence. Suppose that \(u = 11 - x\) represents the
utility of each player, where \(x\) is the number of years of
imprisonment according to each situation. The payoff matrix is as
follows:

\begin{table}[h]
\centering
\caption{Payoff matrix of the example} 
\label{tab:matrix_1} 
\begin{tabular}{rccc}
 & & \multicolumn{2}{c}{Player 2} \\ \cmidrule(lr){3-4}
 & & Cooperates & Defects \\ \cmidrule(r){2-4}
\multirow{2}{*}{Player 1} & Cooperates & $(10, 10)$ & $(1, 11)$ \\
 & Defects & $(11, 1)$ & $(6, 6)$ \\ \bottomrule
\end{tabular}
\end{table}

Let's suppose now that the game has a previous stage: when meeting
before the interrogations, 1 and 2 already know about the dilemma that
will be proposed to them, so they decide to sign the following contract:
the one who confesses will receive an extra six years of sentence. The
matrix becomes the following:

\begin{table}[h]
\centering
\caption{Payoff matrix after the contract} 
\label{tab:matrix_2} 
\begin{tabular}{rccc}
 & & \multicolumn{2}{c}{Player 2} \\ \cmidrule(lr){3-4}
 & & Cooperates & Defects \\ \cmidrule(r){2-4}
\multirow{2}{*}{Player 1} & Cooperates & $(10, 10)$ & $(1, 5)$ \\
 & Defects & $(5, 1)$ & $(0, 0)$ \\ \bottomrule
\end{tabular}
\end{table}

This results in a unique Nash equilibrium, where both parties opt not to
confess, leading to a prison term of only one year, as opposed to the
five years of imprisonment in the original game's unique equilibrium.

This is what we call a \textbf{losing contract}: each player has the option, when
choosing certain strategies, to give up a part of their utility but
never to acquire more or transfer it to another player. One of the characteristics of these contracts is that they are based on
contingencies that are not realized in equilibrium. In our example, the contract penalizes confessing, even though neither player will do so in the final outcome. The single purpose of losing contracts is to
change the equilibrium to eliminate betrayal incentives.

\subsection{General case}

\subsubsection{Two-player games}

Let us take our example to a more general case and study what
mathematical conditions are required to achieve the desired equilibrium
change. Consider the following payoff matrix:

\begin{table}[h]
\centering
\caption{Payoff matrix of the two-player general case} 
\label{tab:matrix_3} 
\begin{tabular}{rccc}
 & & \multicolumn{2}{c}{Player 2} \\ \cmidrule(lr){3-4}
 & & $s_{2,1}$ & $s_{2,2}$ \\ \cmidrule(r){2-4}
\multirow{2}{*}{Player 1} & $s_{1,1}$ & $(u_1(1,1),u_2(1,1))$ & $(u_1(1,2),u_2(1,2))$ \\
 & $s_{1,2}$ & $(u_1(2,1),u_2(2,1))$ & $(u_1(2,2),u_2(2,2))$ \\ \bottomrule
\end{tabular}
\end{table}

For the game to be a prisoner's dilemma (in any classic version, not
just the one in our example), the following conditions must be assumed:

\begin{itemize}
\item
  \(u_{1}(1,2) < u_{1}(2,2) < u_{1}(1,1) < u_{1}(2,1)\),
\item
  \(u_{2}(2,1) < u_{2}(2,2) < u_{2}(1,1) < u_{2}(1,2)\).
\end{itemize}

The number of possible utilities for each player is limited due to the
interrelated nature of these utilities. However, there is no condition
that relates a player's utilities to those of their opponent, at least in
the context of a single game repetition, which is the focus of this
study. Now, let's see what conditions the contract needs to change the
Nash equilibrium so that it remains unique.

Losing contracts are formally defined for two-player games with two
strategies each (we will then provide a definition for \(n\)-player
games where each player \(i\) has \(k_{i}\) possible strategies): a
losing contract is an ordered pair
\(\left( r_{\tilde{k}_{1}},r_{\tilde{k}_{2}} \right)\), where
\(1 \leq \tilde{k}_{i} \leq 2\left( \tilde{k}_{i} \in \mathbb{N} \right)\) for
\(i = 1,2\); where \(r_{\tilde{k}_{i}} \in \mathbb{R}_{\geq 0}\) represents the
utility that player \(i\) gives up by choosing strategy \(s_{i,\tilde{k}_{i}}\).
A losing contract of (0,0) equals the original game, so we assume
\(r_{\tilde{k}_{i}} > 0\) for some \(i \in \{ 1,2\}\) for the contract to make
sense. Let us call
\({\tilde{u}}_{1}\left( \tilde{k}_{1},k \right) = u_{1}\left( \tilde{k}_{1},k \right) - r_{\tilde{k}_{1}}\)
the utility of player 1 if he chooses the strategy \(s_{1,\tilde{k}_{1}}\) and
his opponent chooses the strategy \(s_{2,k}\) in the new game.
Similarly, let us consider
\({\tilde{u}}_{2}\left( k,\tilde{k}_{2} \right) = u_{2}\left( k,\tilde{k}_{2} \right) - r_{\tilde{k}_{2}}\).

Let us note that in our payoff matrix of the general case, because of
how we chose the Nash equilibrium, \(\tilde{k}_{1} = \tilde{k}_{2} = 2\). To simplify notation, we shall denote by $r_i$ the amount that player $i$ forfeits when choosing strategy $s_{i,2}$ for $i=1,2$. (\(s_{1,2},s_{2,1}\)) is the strategy profile that gives the
highest utility to player 1 but the lowest to player 2. Similarly,
(\(s_{1,1},s_{2,2}\)) is the strategy profile that gives the lowest utility to player 1. We conclude that if the equilibrium were to shift to either of these
two strategy profiles, there would be one player who would not be better off
signing the contract.

This is why we consider strategy profile \(\left( s_{1,1},s_{2,1} \right)\):
this set of strategies is the only one that improves the utility of both
participants with respect to \(\left( s_{1,2},s_{2,2} \right)\), i.e.,
with respect to the equilibrium of the original game. Let us determine
the conditions for (\(s_{1,1},s_{2,1}\)) to constitute a new equilibrium
given that now our matrix is as follows:

\begin{table}[h]
\centering
\caption{Payoff matrix after the losing contract} 
\label{tab:matrix_4} 
\begin{tabular}{rccc}
 & & \multicolumn{2}{c}{Player 2} \\ \cmidrule(lr){3-4}
 & & $s_{2,1}$ & $s_{2,2}$ \\ \cmidrule(r){2-4}
\multirow{2}{*}{Player 1} & $s_{1,1}$ & $(u_1(1,1),u_2(1,1))$ & $(u_1(1,2),\tilde{u}_2(1,2))$ \\
 & $s_{1,2}$ & $(\tilde{u}_1(2,1),u_2(2,1))$ & $(\tilde{u}_1(2,2),\tilde{u}_2(2,2))$ \\ \bottomrule
\end{tabular}
\end{table}

We would like this new equilibrium to be the only one. To ensure that
this is the case, we will make \(s_{i,1}\) a dominant strategy for each
player \(i = 1,2\). This implies that
\({\tilde{u}}_{1}(2,k) < u_{1}(1,k)\) $\land$ \({\tilde{u}}_{2}(k,2) < u_{2}(k,1)\)
for each \(j = 1,2\). Therefore, we have:

\begin{itemize}
\item
  \({\tilde{u}}_{1}(2,1) < u_{1}(1,1) \Leftrightarrow u_{1}(2,1) - r_{1} < u_{1}(1,1) \Leftrightarrow r_{1} > u_{1}(2,1) - u_{1}(1,1) > 0\)
  (by initial condition),
\item
  \({\tilde{u}}_{1}(2,2) < u_{1}(1,2) \Leftrightarrow u_{1}(2,2) - r_{1} < u_{1}(1,2) \Leftrightarrow r_{1} > u_{1}(2,2) - u_{1}(1,2) > 0\)
  (by initial condition).
\item
  For both of the above,
  \(r_{1} > {\max}_{k = 1,2}\left\{ u_{1}(2,k) - u_{1}(1,k) \right\}\).
  We can define \(r_{1} =\max_{k = 1,2}\left\{ u_{1}(2,k) - u_{1}(1,k) \right\} + \varepsilon\)
  for \(\varepsilon > 0\).
\item
  Similarly,
  \(r_{2} = {\max}_{k = 1,2}\left\{ u_{2}(k,2) - u_{2}(k,1) \right\} + \varepsilon\).
\end{itemize}

Let us note that, in fact, for the new equilibrium to be unique, it is
sufficient for only one of the following two conditions to be met:
\({\tilde{u}}_{2}(2,2) < u_{2}(2,1)\),
\({\tilde{u}}_{1}(2,2) < u_{1}(1,2)\). Without losing generality,
suppose that 1 finds an \(r_{1}\) that satisfies the first item of the
above but not the second: \({\tilde{u}}_{1}(2,2) \geq u_{1}(1,2)\).
The strategy \(s_{2,1}\) is still dominant for 2, so this \(r_{1}\) does
not bring any benefit to 1. The case for player 2 is similar, so there
are no incentives for players 1 and 2 to choose values for \(r_{1}\) and \(r_{2}\) other than those already
defined.

We can choose \(\varepsilon = 1\) by convention, because as long as it
respects the conditions, the choice of \(\varepsilon\) will not affect
the new equilibrium: we already noted that the contingencies of losing
contracts are never realized in equilibrium, so no one will choose a strategy
entailing a payoff of \(\tilde{u}\) for either player; hence, the
value of \(\varepsilon\) will not modify any payoff of the game. This indicates that no prior choice by the players is necessary before signing the contract: its amounts are already established by definition.

\subsubsection[$n$-player games]{$\bm{n}$-player games}

The definition of losing contracts for $n$-player games with finite
strategies is as follows:

\vspace{1\baselineskip}

\noindent \textbf{Definition 1.} A losing contract for a game of \(n\) players, in which
each player \(i\) has \(k_{i}\) possible strategies, is a set of $n$
finite sequences, where each sequence \(i\) has length \(k_{i} - 1\)
(this definition makes sense as each player has at least two possible
strategies, hence \(k_{i} \geq 2\) \(\forall 1 \leq i \leq n)\). Each
\(i\)-th sequence is defined as
(\(r_{i,\tilde{k}_{1}},\ldots,r_{i,\tilde{k}_{k_{i} - 1}}\)) for
\(1 \leq \tilde{k}_{1},\ldots,\tilde{k}_{k_{i} - 1} \leq k_{i}\left( \tilde{k}_{j} \in \mathbb{N},1\leq j\leq k_i-1 \right)\)
distinct from each other, where \(r_{i,\tilde{k}_{j}} \in \mathbb{R}_{\geq 0}\)
represents the utility given up by player \(i\) when choosing strategy
\(s_{i,\tilde{k}_{j}}\).

\vspace{1\baselineskip}

A losing contract where every sequence is null is equivalent to the
original game; thus, as we did with two players, we assume
\(r_{i,\tilde{k}_{j}} > 0\) for some
\(j \in \left\{ 1,\ldots,k_{i} - 1 \right\}\), at least for some
\(i \in \{ 1,\ldots,n\}\). For ease of notation, it can also be assumed
that \(\forall 1 \leq i \leq n\), the cooperation strategy of each
player is \(s_{i,1}\), and so we can take
\(\tilde{k}_{j} = j + 1\) \(\forall 1 \leq j \leq k_{i} - 1\). Let us define
\({\tilde{u}}_{i}\left( k,s_{- i} \right) = u_{i}\left( k,s_{- i} \right) - r_{i,k}\)
as the utility of player \(i\) if he chooses strategy \(s_{i,k}\)
(with \(2 \leq k \leq k_{i}\)) and the opponents choose \(s_{- i}\) in the
new game.

The solution that will be presented herein is applicable not only to a
general version of the prisoner's dilemma with $n$ players and finite
strategies for each player but also to any $n$-player game in which all
players agree on the equilibrium to be reached, which can be designated
as \textit{cooperative equilibrium}. To this end, the \textit{feasible} regions of each
payoff are defined, as is the concept of optimizing the cooperative
equilibrium. However, first, two remarks are made.

\vspace{1\baselineskip}

\noindent \textbf{Remark 1.} Even though our losing contracts do not permit the exchange
of goods, it can be posited that the utility of the players is
transferable. Prior to the initiation of the game, players have the
capacity to establish coalitions through the signing of alternative
contracts, thereby determining the distribution of goods among coalition
members. The following condition needs to be imposed: if
\(\left( s_{1,1},\ldots,s_{n,1} \right)\) is a cooperative equilibrium,
then every strategy profile
\(P \neq \left( s_{1,1},\ldots,s_{n,1} \right)\) must satisfy
\(\sum_{i = 1}^{n}\mspace{2mu} u_{i}(P) \leq \sum_{i = 1}^{n}\mspace{2mu} u_{i}(1,\ldots,1)\).
If there were a profile \(P \neq \left( s_{1,1},\ldots,s_{n,1} \right)\) such that
\(\sum_{i = 1}^{n}\mspace{2mu} u_{i}(P) > \sum_{i = 1}^{n}\mspace{2mu} u_{i}(1,\ldots,1)\), players could sign a contract to redistribute the gains so that,
\(u_{i}(P) \geq u_{i}(1,\ldots,1)\) \(\forall 1 \leq i \leq n\) with at least
one strict inequality. This would imply that our cooperative
equilibrium is not Pareto-optimal under transferable utility. Therefore,
the above-mentioned condition is tantamount to requiring
Pareto-optimality in our cooperative equilibrium under the assumption
that utility is transferable.

\vspace{1\baselineskip}

\noindent \textbf{Remark 2.} It is found that by imposing this Pareto-optimality
condition in a classical prisoner's dilemma, the following is
required:

\begin{itemize}
\item
  \(u_{1}(2,2) + u_{2}(2,2) \leq u_{1}(1,1) + u_{2}(1,1)\) (it is
  already given to us, since
  \(u_{1}(2,2) < u_{1}(1,1)\) $\land$ \(u_{1}(2,2) < u_{1}(1,1)\) because it
  is a prisoner's Dilemma),
\item
  \(u_{1}(1,2) + u_{2}(1,2) \leq u_{1}(1,1) + u_{2}(1,1)\),
\item
  \(u_{1}(2,1) + u_{2}(2,1) \leq u_{1}(1,1) + u_{2}(1,1)\).
\end{itemize}

The last two are often required (sometimes strictly) as additional
conditions in the classical prisoner's dilemma in the study of repeated
games: if the game is symmetric, they make mutual cooperation preferable
to both players alternating between betraying and being betrayed.

\vspace{1\baselineskip}

\noindent \textbf{Definition 2.} Let \(M\) be a payoff matrix and \(t\) an amount that modifies
\(M\) (\({\tilde{M}}_{t}\) is the matrix modified by \(t\)). Then,
\(t\) is in its feasible region if for each utility
\(u_{i}\left( k,s_{- i} \right)\left( 2 \leq k \leq k_{i} \right)\) of
each player \(i\) that was affected by the amount \(t\),
\(u_{i}\left( k,s_{- i} \right) < u_{i}\left( 1,s_{- i} \right)\) is satisfied in
\({\tilde{M}}_{t}\), and for each utility
\(u_{i}\left( 1,s_{- i} \right)\) of each player \(i\) that was affected too,
\(u_{i}\left( k,s_{- i} \right) < u_{i}\left( 1,s_{- i} \right)\) holds in \(\tilde{M}_t\) \(\forall 2 \leq k \leq k_{i}\).

\vspace{1\baselineskip}

In the case of a contract \(C\) that modifies \(M\), we define
\({\tilde{M}}_{C}\) as the matrix modified by the contract \(C\) (or
\(\tilde{M}\) if it is clear from the context).

\vspace{1\baselineskip}

\noindent \textbf{Definition 3.} Let \(M\) be a payoff matrix and \(C\) a previous contract.
Then, \(C\) optimizes the cooperative equilibrium of the game if the
latter (\(E\)) is a Nash equilibrium in \(\tilde{M}\) and, moreover,
it does not decrease the utility of any of the players at \(E\) with
respect to that obtained in \(M\). That is, \(C\) does not need
to decrease the utility of any player in \(E\) with respect to the
original matrix to make this profile a Nash equilibrium in the new matrix.

\vspace{1\baselineskip}

With these three definitions, we can prove our first result, to which we
will return later.

\vspace{1\baselineskip}

\noindent \textbf{Lemma 1.} \textit{Let \(M\) be the payoff matrix of a game with cooperative
equilibrium and \(C\) a losing contract such that
\(r_{i,k} = {\max}_{{\tilde{s}}_{- i} \in S_{- i}}\left\{ u_{i}\left( k,{\tilde{s}}_{- i} \right) - u_{i}\left( 1,{\tilde{s}}_{- i} \right) \right\} + \varepsilon_{i,k}\)
(with some \(\varepsilon_{i,k} > 0\)) for each
\(1 \leq i \leq n,2 \leq k \leq k_{i}\). Then each amount \(r_{i,k}\) is
in its feasible region and \(C\) optimizes
\(E = \left( s_{1,1},\ldots,s_{n,1} \right)\). Moreover, it makes this
equilibrium the unique equilibrium.}
\smallskip

\noindent \textit{Proof.} Let \(1 \leq i \leq n,2 \leq k \leq k_{i}\) be such that
\(u_{i}\left( k,s_{- i} \right) \geq u_{i}\left( 1,s_{- i} \right)\) at
\(M\) for some fixed \(s_{- i} \in S_{- i}\) (if there are no \(i\) and
\(k\) satisfying this, then \(s_{i,1}\) is a dominant strategy
\(\forall 1 \leq i \leq n\) and no prior contract \(C\) is needed). We
have
\({\tilde{u}}_{i}\left( k,s_{- i} \right) = u_{i}\left( k,s_{- i} \right) - r_{i,k} < u_{i}\left( k,s_{- i} \right) - {\max}_{{\tilde{s}}_{- i} \in S_{- i}}\left\{ u_{i}\left( k,{\tilde{s}}_{- i} \right) - u_{i}\left( 1,{\tilde{s}}_{- i} \right) \right\}\\ \leq u_{i}\left( 1,s_{- i} \right)\),
where the first inequality is satisfied because
\(\varepsilon_{i,k} > 0\), and the se\-cond because, as
\(s_{- i} \in S_{- i}\) is fixed, then
\(u_{i}\left( k,s_{- i} \right) - u_{i}\left( 1,s_{- i} \right) \leq\max_{{\tilde{s}}_{- i} \in S_{- i}}\) \(\left\{ u_{i}\left( k,{\tilde{s}}_{- i} \right) - u_{i}\left( 1,{\tilde{s}}_{- i} \right) \right\}\).
Therefore, \(r_{i,k}\) is in its feasible region. Moreover, since \(i,k,s_{- i}\) such that
\(u_{i}\left( k,s_{- i} \right) \geq u_{i}\left( 1,s_{- i} \right)\) in
\(M\) were arbitrary, it follows that for each
\(1 \leq i \leq n,u_{i}\left( k,s_{- i} \right) < u_{i}\left( 1,s_{- i} \right)\) holds \(\forall 2 \leq k \leq k_{i},s_{- i} \in S_{- i}\)
in \(\tilde{M}\), since the first inequality of the previous chain
was strict and the values in \(\tilde{M}\) are given by  
\({\tilde{u}}_{i}\left( k,s_{- i} \right)\). This tells us that \(s_{i,1}\) is a dominant
strategy \(\forall 1 \leq i \leq n\) in \(\tilde{M}\), which proves
that $E$ is the unique Nash equilibrium in \(\tilde{M}\). In addition,
\(r_{i,k}\) is defined for \(2 \leq k \leq k_{i}\), so \(C\) does not
modify \(u_{i}(E)\) for any \(1 \leq i \leq n\) (these values are the same
in \(M\) and in \(\tilde{M}\)) and therefore \(C\) optimizes \(E\). $\square$

\vspace{1\baselineskip}

We can again take
\(\varepsilon_{i,k} = 1\) \(\forall 1 \leq i \leq n,2 \leq k \leq k_{i}\) by
convention (the reaso\-ning is similar to what we did for two-player
games); thus, the losing contracts are again established by definition
in any finite game with a cooperative equili\-brium. It should be
clarified that beyond the mathematical question, \(\varepsilon_{i,k}\)
could be defined in such a way that the difference between the associated utilities is significant enough for player \(i\) to no longer
choose the strategy \(k\); \(\varepsilon_{i,k} = 1\) is taken assuming
that one unit of utility is the difference needed, although, for
practical purposes, this may vary depending on the situation.

We aim to define the $r_{i,k}$ as in Lemma 1 to ensure a strong Nash equilibrium, thereby eliminating incentives for coalitional deviations, even under transferable utility as assumed in Remark 1. Subsequently, it will be shown that this holds when the game meets certain conditions: the same conditions for which we will define the $n$-player prisoner's dilemma. The approach is based on maintaining the incentives for cooperation within each subset of players, regardless of the strategies chosen by the remaining players.

In the following section, we will show that for the remaining contract types (exchange contracts), there are games satisfying the prisoner's dilemma conditions where all the amounts in these exchange contracts lie within their feasible regions, yet they fail to meet the conclusions of Lemma 1. This indicates that our losing contracts are the only ones satisfying our first lemma, particularly since the prisoner's dilemma is a game with a cooperative equilibrium. At the end of the section, we will also see a prisoner's dilemma for which there is an exchange contract with amounts in its feasible
regions that does not achieve a strong Nash equilibrium.

\vspace{1\baselineskip}

Before proceeding, let us consider an example where defining the $r_{i,k}$ as in Lemma 1 is insufficient to ensure that the cooperative equilibrium is a strong Nash equilibrium. That is, it is not enough for a Nash equilibrium to be unique and also Pareto-optimal under transferable utility: coalitional incentives may still persist. Consider the following three-player game with two strategies each (clearly not a prisoner's dilemma, as a dominant strategy does not even exist for any player), with its payoff matrix divided into two parts:

\vspace{1\baselineskip}

If player 3 chooses \(s_{3,1}\):

\begin{table}[h]
\centering
\caption{Payoff matrix of the example (first part)} 
\label{tab:matrix_5} 
\begin{tabular}{rccc}
 & & \multicolumn{2}{c}{Player 2} \\ \cmidrule(lr){3-4}
 & & $s_{2,1}$ & $s_{2,2}$ \\ \cmidrule(r){2-4}
\multirow{2}{*}{Player 1} & $s_{1,1}$ & $(8,8,8)$ & $(10,3,10)$ \\
 & $s_{1,2}$ & $(3,10,10)$ & $(11,11,1)$ \\ \bottomrule
\end{tabular}
\end{table}

If player 3 chooses \(s_{3,2}\):

\begin{table}[h]
\centering
\caption{Payoff matrix of the example (second part)} 
\label{tab:matrix_6} 
\begin{tabular}{rccc}
 & & \multicolumn{2}{c}{Player 2} \\ \cmidrule(lr){3-4}
 & & $s_{2,1}$ & $s_{2,2}$ \\ \cmidrule(r){2-4}
\multirow{2}{*}{Player 1} & $s_{1,1}$ & $(10,10,3)$ & $(1,11,11)$ \\
 & $s_{1,2}$ & $(11,1,11)$ & $(2,2,2)$ \\ \bottomrule
\end{tabular}
\end{table}

By Remark 1, $(s_{1,1},s_{2,1},s_{3,1})$ must be the cooperative equilibrium of the game, since it is the unique Pareto-optimal profile. Defining $r_{i,2}$ for each $1\leq i\leq3$ as in Lemma 1 with each $\varepsilon_{i,2}=1$ as stated before, we obtain that $r_{1,2}=r_{2,2}=r_{3,2}=2$, and therefore the new matrix $\tilde{M}$ becomes:

\vspace{1\baselineskip}

If player 3 chooses \(s_{3,1}\):

\begin{table}[h]
\centering
\caption{Payoff matrix after the losing contract (first part)} 
\label{tab:matrix_7} 
\begin{tabular}{rccc}
 & & \multicolumn{2}{c}{Player 2} \\ \cmidrule(lr){3-4}
 & & $s_{2,1}$ & $s_{2,2}$ \\ \cmidrule(r){2-4}
\multirow{2}{*}{Player 1} & $s_{1,1}$ & $(8,8,8)$ & $(10,1,10)$ \\
 & $s_{1,2}$ & $(1,10,10)$ & $(9,9,1)$ \\ \bottomrule
\end{tabular}
\end{table}

\clearpage

If player 3 chooses \(s_{3,2}\):

\begin{table}[h]
\centering
\caption{Payoff matrix after the losing contract (second part)} 
\label{tab:matrix_8} 
\begin{tabular}{rccc}
 & & \multicolumn{2}{c}{Player 2} \\ \cmidrule(lr){3-4}
 & & $s_{2,1}$ & $s_{2,2}$ \\ \cmidrule(r){2-4}
\multirow{2}{*}{Player 1} & $s_{1,1}$ & $(10,10,1)$ & $(1,9,9)$ \\
 & $s_{1,2}$ & $(9,1,9)$ & $(0,0,0)$ \\ \bottomrule
\end{tabular}
\end{table}

By Lemma 1 (which can also be verified in the matrix), $(s_{1,1},s_{2,1},s_{3,1})$ is the unique Nash equilibrium in $\tilde{M}$, and the payoffs of this profile remain the same as in $M$; thus, it remains Pareto-optimal. However, notice that any pair of players deviating from $(s_{1,1},s_{2,1},s_{3,1})$ obtains higher payoffs, and therefore the latter is not a strong equilibrium.

Although a formal definition of the $n$-player prisoner's dilemma where each player \(i\) has \(k_{i}\) possible strategies will be
provided later, a three-player prisoner's dilemma where each player has two strategies must be defined for the next section. The following three conditions
will be required:

\begin{itemize}
\item
  \(\forall 1 \leq i \leq 3,u_{i}(2,2,2) < u_{i}(1,1,1)\) (later we will
  prove that it is a consequence of the other two conditions, but for
  now, let's ask for it),
\item
  \(\forall 1 \leq i \leq 3,u_{i}\left( 1,s_{- i} \right) < u_{i}\left( 2,s_{- i} \right)\) \(\forall s_{- i} \in S_{- i}\),
\item
  \(\forall 1 \leq i \leq 3,\forall j \neq i,u_{i}\left( s_{i,2},s_{j,2},s_{k} \right) < u_{i}\left( s_{i,1},s_{j,1},s_{k} \right)\)
  if \(s_{k}\) fixed, \(k \neq i,j\).
\end{itemize}

The first condition shows that joint cooperation is more beneficial to each player than if they all choose to defect. The second states that betrayal is a dominant strategy for all three players. The third one tells us that any two players benefit more if they both cooperate than if they both betray, regardless of the third player's choice; that is, given any fixed strategy by any player, the induced game played by the remaining players is still a prisoner's dilemma, although we will not require the additional conditions of Remark 2 to be met in these induced games. An even weaker condition than that of Remark 1 will be enough.

Finally, two concepts that will be used later will be defined. Consider
a game \(G'\) derived from \(G\) through the signing of a contract.
Given some \(i \in \{ 1,\ldots,n\}\), we will say that signing this
contract is a dominant strategy for \(i\) if there are no incentives for player $i$ to deviate in any coalition in \(G\) with respect to the
worst equilibrium of \(G'\), and
\(u_{i}(s) \leq u_{i}'\left( s' \right)\) is met for any pair of Nash
equilibria \(\left( s,s' \right)\) (\(s\) of \(G\) and \(s'\) strong of
\(G'\)), with at least one strict inequality. If $G$ is a two-player game, we shall say that signing the contract is a dominant strategy for $i$ regardless of the condition on the coalitional incentives of $i$ in $G$, and without requiring $s'$ to be strong in $G'$, but simply Nash equilibria. In addition, if there is a contract $C$ such that signing it is dominant for all players, we say that $G'$ (derived from $G$ by $C$) is a resolution of $G$, or that $G$ is \textbf{solved by contract} $C$ (or that it is solved by $G'$).

\subsection{Discussion}

Losing contracts are agreements between game parties; however, for the
contracts to be binding, something more is required than is the case
with exchange contracts: the latter require an external agent with the
authority to enforce the exchange of goods if necessary, while in losing
contracts, even if the external agent in question does not impose
sanctions, it must be able to collect the goods that the players give
up or at least be able to ensure that the goods are destroyed and not
retained by any other player.

Moreover, so far, we have already achieved two advantages: the first is
that in losing contracts, each player's amounts affect only his own
utility, which gives us more control over the equilibrium in games with
more than two players; and the second is that our contracts are
established by definition (beyond the \(\varepsilon_{i,k}\), which do
not affect the utilities of the others). This transformation effectively
reduces the pre-game stage to a binary decision for each player, namely,
whether to sign the contract or not. Furthermore, if the cooperative
equilibrium is strong and provides them more utility than the Nash equilibria, and there are
no coalitional incentives in $G$, then signing the contract is a dominant
strategy for all players and \(G\) is solved by the losing contract.

\section{Disadvantages of exchanging}

\subsection{Rewarding cooperation}

First, an example of a prisoner's dilemma is given to see what happens.
Consider the following matrix \(M\):

\begin{table}[h]
\centering
\caption{Payoff matrix of the example} 
\label{tab:matrix_9} 
\begin{tabular}{rccc}
 & & \multicolumn{2}{c}{Player 2} \\ \cmidrule(lr){3-4}
 & & $s_{2,1}$ & $s_{2,2}$ \\ \cmidrule(r){2-4}
\multirow{2}{*}{Player 1} & $s_{1,1}$ & $(7, 5)$ & $(2, 9)$ \\
 & $s_{1,2}$ & $(9, 1)$ & $(3, 2)$ \\ \bottomrule
\end{tabular}
\end{table}

Let us call \(p_{i}\) the amount that player \(i\) offers to pay in the
pre-contract stage to the other player in case the latter chooses the
strategy \(s_{j,1}\) with \(j \neq i\). The new matrix \(\tilde{M}\)
leaves us with the following:

\begin{table}[h]
\centering
\caption{Payoff matrix after the contract} 
\label{tab:matrix_10} 
\begin{tabular}{rccc}
 & & \multicolumn{2}{c}{Player 2} \\ \cmidrule(lr){3-4}
 & & $s_{2,1}$ & $s_{2,2}$ \\ \cmidrule(r){2-4}
\multirow{2}{*}{Player 1} & $s_{1,1}$ & $(7+p_2-p_1, 5+p_1-p_2)$ & $(2+p_2, 9-p_2)$ \\
 & $s_{1,2}$ & $(9-p_1, 1+p_1)$ & $(3, 2)$ \\ \bottomrule
\end{tabular}
\end{table}

For (\(s_{1,1},s_{2,1}\)) to be a Nash equilibrium, the following must
be met:

\begin{itemize}
\item
  \(9 - p_{1} < 7 + p_{2} - p_{1} \Leftrightarrow 2 < p_{2}\),
\item
  \(9 - p_{2} < 5 + p_{1} - p_{2} \Leftrightarrow 4 < p_{1}\).
\end{itemize}

Additionally, in this case, notice that if
\(p_{1} > 4\) $\land$ \(p_{2} > 2\), then $p_1$ and $p_2$ are in their respective feasible regions, which implies that
(\(s_{1,1},s_{2,1}\)) is the unique equilibrium. Suppose 2 is willing to
pay \(p_{2} = 3\). We know that \(p_{1} = 4 + \varepsilon\) for some
\(\varepsilon > 0\), so matrix \(\tilde{M}\) becomes:

\begin{table}[h]
\centering
\caption{Payoff matrix if $p_2=3$} 
\label{tab:matrix_11} 
\begin{tabular}{rccc}
 & & \multicolumn{2}{c}{Player 2} \\ \cmidrule(lr){3-4}
 & & $s_{2,1}$ & $s_{2,2}$ \\ \cmidrule(r){2-4}
\multirow{2}{*}{Player 1} & $s_{1,1}$ & $(6-\varepsilon, 6+\varepsilon)$ & $(5, 6)$ \\
 & $s_{1,2}$ & $(5-\varepsilon, 5+\varepsilon)$ & $(3, 2)$ \\ \bottomrule
\end{tabular}
\end{table}

We can see that, in the new equilibrium (\(s_{1,1},s_{2,1}\)), player 1
obtains a utility of \(u_{1}(1,1) < 6\), while in the previous game this
value was \(u_{1}(1,1) =\) 7. That is, in the new Nash equilibrium at
\(\tilde{M}\) (choosing an \(\varepsilon < 3\)), 1 obtains more than
in the original equilibrium (\(u_{1}(2,2) = 3\)) but less than in the
strategy profile (\(s_{1,1},s_{2,1}\)) of the original \(M\) matrix game for
whichever \(\varepsilon\) is chosen.

This means that one player (in this case player 1), despite having
incentives to sign the contract, is forced to give up a part of his
original utility to achieve cooperative equilibrium: the other player
(in this case player 2) has the chance to increase his utility by
choosing a \(p_{2} > 2\) that ends up damaging his opponent with respect
to cooperation at \(M\). It is thus concluded that contracts rewarding
cooperation may not optimize the cooperative equilibrium.

Even worse, let us note that if \(u_{1}(2,2) = 6\) at \(M\), player 1
obtains a lower utility at \(\left( s_{1,1},s_{2,1} \right)\) from
\(\tilde{M}\) than in equilibrium from \(M\); thus, if player 2
offers an amount \(p_{2} = 3\), it is not even in 1's interest to sign
the contract.

\subsection{Punishing defection}

\subsubsection{Conditional amounts}

Contracts punishing defection always optimize the Nash equilibrium. In
addition, they always make the cooperative equilibrium unique if there are only two players. Let us see an example with three players where, after signing the contract, the cooperative
equilibrium is not the unique Nash equilibrium of the payoffs matrix $\tilde{M}$.

In these contracts, players must offer an amount to be shared among the other two in the event of betrayal. We will assume that they can choose how they prefer to distribute this amount in each contingency, under the condition that they must distribute it in full. In the following example, we assume that the three players agree on one aspect of their choice: in the event that two players choose to betray and one chooses to cooperate, regardless of their identity, the two betrayers will pay each other their full amount. This is done to prevent the cooperator from gaining too much advantage by being the only one who does not have to pay.

As there are three players, let's consider the following game with its
payoff matrix \(M\) divided into two parts:

\vspace{1\baselineskip}

If player 3 cooperates (chooses \(s_{3,1}\)):

\begin{table}[h]
\centering
\caption{Payoff matrix of the example (first part)} 
\label{tab:matrix_12} 
\begin{tabular}{rccc}
 & & \multicolumn{2}{c}{Player 2} \\ \cmidrule(lr){3-4}
 & & $s_{2,1}$ & $s_{2,2}$ \\ \cmidrule(r){2-4}
\multirow{2}{*}{Player 1} & $s_{1,1}$ & $(7,7,7)$ & $(4,8,4)$ \\
 & $s_{1,2}$ & $(8,5,3)$ & $(5,6,1)$ \\ \bottomrule
\end{tabular}
\end{table}

If player 3 defects (chooses \(s_{3,2}\)):

\begin{table}[h]
\centering
\caption{Payoff matrix of the example (second part)} 
\label{tab:matrix_13} 
\begin{tabular}{rccc}
 & & \multicolumn{2}{c}{Player 2} \\ \cmidrule(lr){3-4}
 & & $s_{2,1}$ & $s_{2,2}$ \\ \cmidrule(r){2-4}
\multirow{2}{*}{Player 1} & $s_{1,1}$ & $(5,5,8)$ & $(2,6,5)$ \\
 & $s_{1,2}$ & $(6,3,4)$ & $(3,4,2)$ \\ \bottomrule
\end{tabular}
\end{table}

First, let's see that this game is a prisoner's dilemma for three
players with two strategies each. For this, the three conditions mentioned at the end of the
previous section must be met:

\begin{itemize}
\item
  \(\forall 1 \leq i \leq 3,u_{i}(2,2,2) < u_{i}(1,1,1)\),
\item
  \(\forall 1 \leq i \leq 3,u_{i}\left( 1,s_{- i} \right) < u_{i}\left( 2,s_{- i} \right)\forall\) \(s_{- i} \in S_{- i}\),
\item
  \(\forall 1 \leq i \leq 3,\forall j \neq i,u_{i}\left( s_{i,2},s_{j,2},s_{k} \right) < u_{i}\left( s_{i,1},s_{j,1},s_{k} \right)\)
  if \(s_{k}\) fixed, \(k \neq i,j\).
\end{itemize}

The first condition is met. Let's analyze the second one for each player:

\begin{itemize}
\item
  \(u_{1}(1,2,2) = 2 < u_{1}(2,2,2) = 3,u_{1}(1,2,1) = 4 < u_{1}(2,2,1) = 5\),\\
  \(u_{1}(1,1,2) = 5 < u_{1}(2,1,2) = 6,u_{1}(1,1,1) = 7 < u_{1}(2,1,1) = 8\),
\item
  \(u_{2}(2,1,2) = 3 < u_{2}(2,2,2) = 4,u_{2}(2,1,1) = 5 < u_{2}(2,2,1) = 6\),\\
  \(u_{2}(1,1,2) = 5 < u_{2}(1,2,2) = 6,u_{2}(1,1,1) = 7 < u_{2}(1,2,1) = 8\),
\item
  \(u_{3}(2,2,1) = 1 < u_{3}(2,2,2) = 2,u_{3}(2,1,1) = 3 < u_{3}(2,1,2) = 4\),\\
  \(u_{3}(1,2,1) = 4 < u_{3}(1,2,2) = 5,u_{3}(1,1,1) = 7 < u_{3}(1,1,2) = 8\).
\end{itemize}

Let's do the same with the third one:

\begin{itemize}
\item
  \(u_{1}(2,2,2) = 3 < u_{1}(1,1,2) = 5,u_{1}(2,2,1) = 5 < u_{1}(1,1,1) = 7\),\\
  \(u_{1}(2,2,2) = 3 < u_{1}(1,2,1) = 4,u_{1}(2,1,2) = 6 < u_{1}(1,1,1) = 7\),
\item
  \(u_{2}(2,2,2) = 4 < u_{2}(2,1,1) = 5,u_{2}(1,2,2) = 6 < u_{2}(1,1,1) = 7\),\\
  \(u_{2}(2,2,2) = 4 < u_{2}(1,1,2) = 5,u_{2}(2,2,1) = 6 < u_{2}(1,1,1) = 7\),
\item
  \(u_{3}(2,2,2) = 2 < u_{3}(1,2,1) = 4,u_{3}(2,1,2) = 4 < u_{3}(1,1,1) = 7\),\\
  \(u_{3}(2,2,2) = 2 < u_{3}(2,1,1) = 3,u_{3}(1,2,2) = 5 < u_{3}(1,1,1) = 7\).
\end{itemize}

We see that all of them are satisfied, and, therefore, we are in a version
of the prisoner's dilemma. Similar to the contracts rewarding
cooperation, now we will call \(r_{i}\) the amount that the player
\(i\) offers to pay to the other players each time he betrays. We
should recall that in our example, in cases where two players betray
and one cooperates, the betrayers choose to pay each other their full
amounts and give nothing to the cooperator. Let us also assume that in
all other cases, player 1 prefers to give his entire amount to
player 2, player 2 prefers to give it to player 3, and player 3 to
player 1.

Before presenting the new payoff matrix, let us note that due to the
inequa\-lities in the second condition, if
\(r_{i} > 1\) \(\forall 1 \leq i \leq 3 \Rightarrow \left( s_{1,1},s_{2,1},s_{3,1} \right)\)
is a Nash equilibrium, since if this occurs, given
\(s_{- i} = (1,1),u_{i}\left( s_{i,2},s_{- i} \right) - r_{i} < 8 - 1 = 7 = u_{i}(1,1,1)\) \(\forall 1 \leq i \leq 3\). It can be seen more clearly in the new matrix \(\tilde{M}\):

\vspace{1\baselineskip}

If player 3 cooperates (chooses \(s_{3,1}\)):

\begin{table}[h]
\centering
\caption{Payoff matrix after the contract (first part)} 
\label{tab:matrix_14} 
\begin{tabular}{rccc}
 & & \multicolumn{2}{c}{Player 2} \\ \cmidrule(lr){3-4}
 & & $s_{2,1}$ & $s_{2,2}$ \\ \cmidrule(r){2-4}
\multirow{2}{*}{Player 1} & $s_{1,1}$ & $(7,7,7)$ & $(4,8-r_2,4+r_2)$ \\
 & $s_{1,2}$ & $(8-r_1,5+r_1,3)$ & $(5-r_1+r_2,6-r_2+r_1,1)$ \\ \bottomrule
\end{tabular}
\end{table}

If player 3 defects (chooses \(s_{3,2}\)):

\begin{table}[h]
\centering
\caption{Payoff matrix after the contract (second part)} 
\label{tab:matrix_15} 
\resizebox{\linewidth}{!}{
\begin{tabular}{rccc}
 & & \multicolumn{2}{c}{Player 2} \\ \cmidrule(lr){3-4}
 & & $s_{2,1}$ & $s_{2,2}$ \\ \cmidrule(r){2-4}
\multirow{2}{*}{Player 1} & $s_{1,1}$ & $(5+r_3,5,8-r_3)$ & $(2,6-r_2+r_3,5-r_3+r_2)$ \\
 & $s_{1,2}$ & $(6-r_1+r_3,3,4-r_3+r_1)$ & $(3-r_1+r_3,4-r_2+r_1,2-r_3+r_2)$ \\ \bottomrule
\end{tabular}
}
\end{table}

Let us note that the condition we mentioned
before: \(r_{i} > 1\) \(\forall 1 \leq i \leq 3\); is the only one
(necessary and sufficient) for (\(s_{1,1},s_{2,1},s_{3,1}\)) to be a
Nash equilibrium; however, for it to be the only one, we need at least
one of the following conditions to also be met, so
\(\left( s_{1,2},s_{2,2},s_{3,2} \right)\) is not an equilibrium:

\begin{itemize}
\item
  \(u_{1}(2,2,2) = 3 - r_{1} + r_{3} < u_{1}(1,2,2) = 2 \Leftrightarrow 1 + r_{3} < r_{1}\),
\item
  \(u_{2}(2,2,2) = 4 - r_{2} + r_{1} < u_{2}(2,1,2) = 3 \Leftrightarrow 1 + r_{1} < r_{2}\),
\item
  \(u_{3}(2,2,2) = 2 - r_{3} + r_{2} < u_{3}(2,2,1) = 1 \Leftrightarrow 1 + r_{2} < r_{3}\).
\end{itemize}

That is, one of the players must be willing to offer an amount exceeding
the amount of another player by more than one unit. However, there is no
guarantee that this will happen. Specifically, if
\(r: = r_{1} = r_{2} = r_{3}\) (e.g., \(r = 2\)), the matrix
\(\tilde{M}\) looks like this:

\vspace{1\baselineskip}

If player 3 cooperates (chooses \(s_{3,1}\)):

\enlargethispage{1\baselineskip}
\begin{table}[H]
\centering
\caption{Payoff matrix if $r_1=r_2=r_3=2$ (first part)} 
\label{tab:matrix_16} 
\begin{tabular}{rccc}
 & & \multicolumn{2}{c}{Player 2} \\ \cmidrule(lr){3-4}
 & & $s_{2,1}$ & $s_{2,2}$ \\ \cmidrule(r){2-4}
\multirow{2}{*}{Player 1} & $s_{1,1}$ & $(7,7,7)$ & $(4,6,6)$ \\
 & $s_{1,2}$ & $(6,7,3)$ & $(5,6,1)$ \\ \bottomrule
\end{tabular}
\end{table}

If player 3 defects (chooses \(s_{3,2}\)):

\begin{table}[h]
\centering
\caption{Payoff matrix if $r_1=r_2=r_3=2$ (second part)} 
\label{tab:matrix_17} 
\begin{tabular}{rccc}
 & & \multicolumn{2}{c}{Player 2} \\ \cmidrule(lr){3-4}
 & & $s_{2,1}$ & $s_{2,2}$ \\ \cmidrule(r){2-4}
\multirow{2}{*}{Player 1} & $s_{1,1}$ & $(7,5,6)$ & $(2,6,5)$ \\
 & $s_{1,2}$ & $(6,3,4)$ & $(3,4,2)$ \\ \bottomrule
\end{tabular}
\end{table}

These choices of \(r_{1},r_{2},r_{3} > 1\) clearly show that there are
amounts satisfying the conditions for (\(s_{1,1},s_{2,1},s_{3,1}\)) to be a Nash equilibrium, but they do not make it unique, as in
this case, \(\left( s_{1,2},s_{2,2},s_{3,2} \right)\) is also an
equilibrium. This is because the last three conditions we mentioned
depend on two different amounts rather than just one, which means that
choosing any single amount cannot guarantee that any of these conditions
will hold: they also depend on the choice of another amount made by
another player, and, in principle, there is no way to force them to
hold. It is thus concluded that our example illustrates that contracts
punishing betrayal may not determine a unique equilibrium, even when payoffs lie in their respective feasible regions.

It should be clarified that if they realized this, the three players
could jointly choose three amounts that meet any of the conditions
for the cooperative equilibrium to be unique, although
it would entail an instance of pre-contract bargaining. Moreover, as
indicated before, it might be the case that no player is willing to offer an
amount exceeding another by more than one unit: the fairest situation
would be for all three conditions to be met, but this would have to be
\(r_{3} < r_{1} < r_{2} < r_{3}\), which is absurd.

\subsubsection{Unconditional amounts}

In the aforementioned example, the amounts of each player are not only
conditioned according to their strategy but also in consideration of the
choices made by the other players. Suppose now that this is not allowed
and that each player \(i\) can choose how to distribute the total amount
according to their strategy but without their choice depending on the
strategies chosen by the others.

To accomplish this objective, it is necessary to examine the following
example, which is analogous to the previous one and which employs the
matrix of the game \(M\) divided into two parts:

\vspace{1\baselineskip}

If player 3 cooperates (chooses \(s_{3,1}\)):

\enlargethispage{2\baselineskip}
\begin{table}[H]
\centering
\caption{Payoff matrix of the example (first part)} 
\label{tab:matrix_18} 
\begin{tabular}{rccc}
 & & \multicolumn{2}{c}{Player 2} \\ \cmidrule(lr){3-4}
 & & $s_{2,1}$ & $s_{2,2}$ \\ \cmidrule(r){2-4}
\multirow{2}{*}{Player 1} & $s_{1,1}$ & $(7,7,11)$ & $(4,8,4)$ \\
 & $s_{1,2}$ & $(16,5,3)$ & $(5,6,1)$ \\ \bottomrule
\end{tabular}
\end{table}

If player 3 defects (chooses \(s_{3,2}\)):

\begin{table}[h]
\centering
\caption{Payoff matrix of the example (second part)} 
\label{tab:matrix_19} 
\begin{tabular}{rccc}
 & & \multicolumn{2}{c}{Player 2} \\ \cmidrule(lr){3-4}
 & & $s_{2,1}$ & $s_{2,2}$ \\ \cmidrule(r){2-4}
\multirow{2}{*}{Player 1} & $s_{1,1}$ & $(5,5,12)$ & $(2,6,5)$ \\
 & $s_{1,2}$ & $(6,3,4)$ & $(3,4,2)$ \\ \bottomrule
\end{tabular}
\end{table}

We only changed the values of \(u_{1}(2,1,1),u_{3}(1,1,1)\) and
\(u_{3}(1,1,2)\) with respect to the previous matrix; thus, to see that
this game is a prisoner's dilemma, it is enough to examine the
conditions under which these values appear:

\begin{itemize}
\item
  \(u_{1}(1,1,1) = 7 < u_{1}(2,1,1) = 16\),
\item
  \(u_{3}(1,1,1) = 11 < u_{3}(1,1,2) = 12\),
\item
  \(u_{3}(2,1,2) = 4 < u_{3}(1,1,1) = 11\),
\item
  \(u_{3}(1,2,2) = 5 < u_{3}(1,1,1) = 11\).
\end{itemize}

It is sufficient to compare (\(s_{1,1},s_{2,1},s_{3,1}\)) with
(\(s_{1,2},s_{2,1},s_{3,1}\)) to prove that contracts punishing betrayal
can prevent the cooperative equilibrium from being a strong Nash equilibrium if utility is transferable. To this end, the following scenarios
will be analyzed: in the first one, each player can choose how to split
his amount each time he betrays, and in the second and more restrictive
one, each player must split his total amount equally between the other
two players, giving half to each one.

Before starting, it is necessary to consider that no coalition is
possible: if player 1 wants to offer 2 and 3 part of his utility in
order to make them cooperate, he must offer at least 2 to player 2 and 8
to player 3 to equal the utility they would obtain in the cooperative
equilibrium. However, if he offers them those two amounts, he himself
keeps \(16 - 2 - 8 = 6\) utility, which is less than what he would get
in a cooperative equilibrium, so he cannot encourage others to choose to cooperate without harming
himself. This is because Remark 1 is met for the cooperative
equilibrium
\(\left( s_{1,1},s_{2,1},s_{3,1} \right):\sum_{i = 1}^{3}\mspace{2mu} u_{i}(2,1,1) = 16 + 5 + 3 = 24 < \sum_{i = 1}^{3}\mspace{2mu} u_{i}(1,1,1) = 7 + 7 + 11 = 25\),
and it is easy to see that inequality also occurs for the rest of the
strategy profiles. Now, let's analyze the two cases.

In the first one, we suppose that player 1 prefers to give the whole
amount to player 2, player 2 prefers to give it to player 3, and player
3 to player 1. The values of the strategy profiles that matter to us in the
new matrix \(\tilde{M}\) are as follows:

\begin{itemize}
\item 
    $\left( s_{1,1},s_{2,1},s_{3,1} \right) = (7,7,11),\left( s_{1,2},s_{2,1},s_{3,1} \right) = \left( 16 - r_{1},5 + r_{1},3 \right)$
\end{itemize}

For (\(s_{1,1},s_{2,1},s_{3,1}\)) to be a Nash equilibrium, it must be
the case that \(16 - r_{1} < 7 \Leftrightarrow 9 < r_{1}\). Suppose 1
chooses \(r_{1} = 10\). In the new game,
\(\left( s_{1,2},s_{2,1},s_{3,1} \right) = (6,15,3)\). Now, player 2,
seeing that \(s_{3,1}\) is dominant for 3 in the matrix
\(\tilde{M}\), can offer a payoff \(p\) \((1 < p < 8)\) to 1 to not
cooperate, which leaves them both with a higher utility than what they
would obtain in the cooperative equilibrium, since for
\(1:\ 6 + p > 7 = u_{1}(1,1,1)\) and for 2:
\(15 - p > 15 - 8 = 7 = u_{2}(1,1,1)\). Then, as noted earlier,
considering transferable utility, we find a possible contract under
which (\(s_{1,1},s_{2,1},s_{3,1}\)) is not a strong Nash equilibrium, as
there are incentives for a coalition to deviate.

The analysis in the second scenario is similar: the values of the
assignments in the new matrix are the same and the condition for
\(r_{1}\) is the same. Suppose player 1 chooses \(r_{1} = 10\) again.
Then, in this new matrix $\tilde{M}$,
\(\left( s_{1,2},s_{2,1},s_{3,1} \right) = \left( 16 - r_{1},5 + \frac{r_{1}}{2},3 + \frac{r_{1}}{2} \right) = (6,10,8)\).
As in the previous example, player 2 can take advantage of the fact that
\(s_{3,1}\) is dominant for 3 and offer a payoff $p$ $(1<$ $p<3)$ to player 1 in exchange for deviating to not cooperate. For instance, if $p=2$, both players obtain a higher utility than in the cooperative equilibrium: player 1 receives: \(6 + 2 = 8 > 7 = u_{1}(1,1,1)\), while player 2 gets:
\(10 - 2 = 8 > 7 = u_{2}(1,1,1)\). Thus, not even under the constraint
that players must split their amounts equally among the others can we
ensure that the cooperative equilibrium of the new matrix
\(\tilde{M}\) is strong under contracts punishing
betrayal, even when the amounts are in their feasible regions.

This last example illustrates that coalition incentives can arise in exchange contracts. The problem stems from the fact that the amount paid may remain within a subset of players who, as shown, may re-offer a portion to the original owner. This breaks the strong equilibrium if the amount is high enough that the coalition members would rather split it than receive their respective payoffs in the cooperative equilibrium of the game.

\section{The Prisoner's Dilemma}

\subsection{Motivation}

In this section we consider situations in which players may lose the ability to make decisions. This could be due to the impossibility of choosing certain strategies because of restrictions external to the game, such as budgetary or regulatory constraints, or any other restriction limiting the players' available resources. Specifically, we will focus on games whose restricted versions are prisoner's dilemmas. This approach allows us to maintain the tension between individual interest and collective benefit without requiring that all players have the possibility to cooperate.

Assuming these restrictions, the following problem arises: signing the contract may no longer be a dominant strategy, as players lack the guarantee that they will be able to choose their cooperative strategies later on. One way to resolve this is by reducing the penalty amount whenever a player obtains lower utility in the new game than in the original matrix equilibrium (setting it to zero if necessary). We provide a method to reduce this amount such that signing the contract remains dominant, while ensuring that cooperation also remains dominant during the game.

Our objective is to design a contract in which cooperation profiles constitute a strong equilibrium in every restricted game, thereby preserving each player's incentives to cooperate despite potential restrictions on others. We will then define a weaker version representing a public goods game, which requires a minimum level of collective cooperation to maintain the dilemma. The contract will also prove useful in this latter weak version.

\subsection{Strong equilibrium}

Our definition for an $n$-player prisoner's dilemma has already been given
and is thus formalized with the following two definitions:

\vspace{1\baselineskip}

\noindent \textbf{Definition 4.} Let \(G\) be a two-player game, where player 1 has
\(k_{1}\) strategies and player 2 has \(k_{2}\) strategies. \(G\) is a prisoner's
dilemma if every pair of contiguous strategies
\(\left( s_{1,k},s_{1,k + 1} \right)\) of
\(1\) (with \(1 \leq k < k_{1} \)) and
\(\left( s_{2,\tilde{k}},s_{2,\tilde{k} + 1}\right)\)
of 2 (with \(1 \leq \tilde{k} < k_{2}\)) fulfill that game $ G_{k,\tilde{k}} := \left\langle \{ 1,2\}, \left(\left\{ s_{1,k},s_{1,k + 1} \right\},\left\{ s_{2,\tilde{k}},s_{2,\tilde{k} + 1} \right\} \right)\right\rangle$
is a classical prisoner's dilemma. By convention, we will arrange
\(1,\ldots,k_{i}\) (for \(i = 1,2\)) with respect to the degree of
cooperation in a decreasing order (\(s_{i,1}\) the most cooperative and
\(s_{i,k_{i}}\) the least for \(i = 1,2\)).

\vspace{1\baselineskip}

\noindent \textbf{Definition 5.} Let \(G\) be a game of $n$ players with finite strategies each.
\(G\) is a prisoner's dilemma if for every
\(A \subseteq \{ 1,\ldots,n\}\) (henceforth assumed to be non-empty) such that \(|A| \leq n - 2\), it holds
that \(\forall s_{A} \in S_{A}\), \(G_{- A}\left( s_{A} \right)\) is a
prisoner's dilemma. That is, any game induced by any fixed
strategy profile of a subset of players with cardinality less than or equal to \(n - 2\), is a prisoner's dilemma.

\vspace{1\baselineskip}

The two-player prisoner's dilemma with finite strategies each is
thus well defined. Let us show that the $n$-player version is also well
defined for any \(n \geq 3\).

\vspace{1\baselineskip}

\noindent \textbf{Lemma 2.} \textit{The $n$-player prisoner's dilemma with finite strategy sets
is well-defined.}
\smallskip

\noindent \textit{Proof.} First consider an $n$-player prisoner's dilemma in which each player
has two strategies. Suppose \(n \geq 3\) and any prisoner's dilemma of $m$
players \((m \leq n - 1)\) with two strategies each is well defined. Let
\(A \subseteq \{ 1,\ldots,n\}\) such that
\(|A| \leq n - 2\) and \(s_{A} \in S_{A}\). By our Definition 5,
\(G_{- A}\left( s_{A} \right)\) must be a prisoner's dilemma, but this
is well defined by hypothesis, since
\(2 \leq \left| G_{- A}\left( s_{A} \right) \right| \leq n - 1\). By
induction on $n$, a prisoner's dilemma of $n$ players with two strategies
each is well defined \(\forall n \in \mathbb{N}\). It is imperative to
prove that this outcome remains valid for any given game in which players
possess a finite number of strategies. We can start by proving that it
is valid if a single player has more than two strategies. Without loss
of generality, we can assume that player 1 has \(k_{1} \geq 3\)
strategies and that our result holds for all $m$-player prisoner's dilemma
(with \(m \leq n - 1\)). Let \(A \subseteq \{ 1,\ldots,n\}\) such that
\(|A| \leq n - 2\) and \(s_{A} \in S_{A}\). If
\(1 \in A\), \(G_{- A}\left( s_{A} \right)\) is well defined by the result we
tested before, as it is a prisoner's dilemma where each player has two
strategies. Moreover, if \(1 \notin A\), \(G_{- A}\left( s_{A} \right)\) is also
well defined by hypothesis, since
\(2 \leq \left| G_{- A}\left( s_{A} \right) \right| \leq n - 1\). By
induction on $n$, an $n$-player prisoner's dilemma in which one player has more
than two strategies is well defined \(\forall n \in \mathbb{N}\). Before
testing the full proposal, we can see that it is valid if two players
have more than two strategies each. Now, without loss of generality, let
us assume that \(k_{1},k_{2} \geq 3\) and that our result holds for any
$m$-player prisoner's dilemma (\(m \leq n - 1\)). Let
\(A \subseteq \{ 1,\ldots,n\}\) such that
\(|A| \leq n - 2\) and \(s_{A} \in S_{A}\). If
\(1 \in A\) \(\vee\) \(2 \in A\), \(G_{- A}\left( s_{A} \right)\) is well defined, as it
has one or no players with more than two strategies. Moreover, if
\(1,2 \notin A\), \(G_{- A}\left( s_{A} \right)\) is also well defined by
hypothesis, since
\(2 \leq \left| G_{- A}\left( s_{A} \right) \right| \leq n - 1\). By
induction on $n$, it is valid for two players with more than two
strategies each \(\forall n \in \mathbb{N}\). Finally, let us induce the
number of players who have more than two strategies. We determine
\(3 \leq m \leq n\) and assume that any prisoner's dilemma of
\(\tilde{n}\) players (with \(m \leq \tilde{n}\)) in which players
with more than two strategies are less than \(m\) is well defined.
Suppose also that if \(\tilde{n} < n\), it is valid if the players
with more than two strategies are exactly $m$. Note that we have already
seen that for \(m = 2\), the two hypotheses are fulfilled, so we already
have our base case. Without loss of generality, let us assume
\(k_{1},\ldots,k_{m} \geq 3\). Let \(A \subseteq \{ 1,\ldots,n\}\) such that
\(|A| \leq n - 2\) and \(s_{A} \in S_{A}\). If \(\exists i \in A\) such that
\(1 \leq i \leq m\), \(G_{- A}\left( s_{A} \right)\) is well defined, since
it has less than $m$ players with more than two strategies each (first
hypothesis). Moreover, if
\(i \notin A\) \(\forall 1 \leq i \leq m\), \(G_{- A}\left( s_{A} \right)\) is
also well defined, as it is a game of \(\tilde{n} < n\) players of
which $m$ have more than two strategies (second hypothesis). Therefore, by
induction on $m$, the result is valid for any number of players with more
than two strategies. $\square$

\vspace{1\baselineskip}

Before continuing, as intuitive as it may be, let us prove the following
result that will be useful in several demonstrations.

\vspace{1\baselineskip}

\noindent \textbf{Lemma 3.} \textit{In a prisoner's dilemma, if some (non-empty) subset of players
change their strategies to more cooperative ones and the rest keep the
same strategies, the latter obtain a higher utility.}
\smallskip

\noindent \textit{Proof.} Suppose there are $n$ players and each player \(i\) has \(k_{i}\)
strategies. Furthermore, without loss of generality, let
\(A = \{ m,\ldots,n\}\) (with \(2 \leq m \leq n)\) and suppose that
each \(i \in A\) chooses the strategy \(s_{i,{\tilde{k}}_{i}}\) at
\({\tilde{s}}_{A}\) and \(s_{i,k_{i}^{*}}\) at \(s_{A}^{*}\) such
that \({\tilde{k}}_{i} > k_{i}^{*}\) and that 1 chooses the strategy
\(s_{1,k}\) at \(s_{- A}\) (fixed). Let us consider the induced game
\(G_{- \{ 2,\ldots,n\} \setminus \{ m\}}\left( s_{- A} \setminus \left\{ s_{1,k} \right\},{\tilde{s}}_{A} \setminus \left\{ s_{m,{\tilde{k}}_{m}} \right\} \right)\).
If \(k > 1\), we have that due to the joint cooperation of 1 and
\(m,u_{1}\left( s_{- A},{\tilde{s}}_{A} \right) < u_{1}\left( k - 1,s_{- A} \setminus \left\{ s_{1,k} \right\},s_{m,\tilde{k}_m - 1},s_{m + 1,{\tilde{k}}_{m + 1}},\ldots,s_{n,{\tilde{k}}_{n}} \right)\).
In addition, as \(s_{1,k}\) do\-minates
$s_{1,k - 1}$, $u_{1}\left( k - 1,s_{- A} \setminus\left\{ s_{1,k} \right\},s_{m,{\tilde{k}}_m - 1},s_{m + 1,{\tilde{k}}_{m + 1}},\ldots,s_{n,{\tilde{k}}_{n}} \right) < u_{1}\Big(s$ $_{- A},s_{m,{\tilde{k}}_m - 1},s_{m + 1,{\tilde{k}}_{m + 1}},\ldots,s_{n,{\tilde{k}}_{n}} \Big) \Rightarrow u_{1}\left( s_{- A},{\tilde{s}}_{A} \right) < u_{1}\Big( s_{- A},s_{m,{\tilde{k}}_{m} - 1},s_{m + 1},$ $_{\tilde{k}_{m + 1}},\ldots,s_{n,{\tilde{k}}_{n}} \Big)$.
Therefore, by inductive reasoning,
$u_{1}\left( s_{- A},{\tilde{s}}_{A}\right) < u_{1}\Big( s_{- A},$ $s_{m,k_{m}^{*}},s_{m + 1,{\tilde{k}}_{m + 1}},\ldots,s_{n,{\tilde{k}}_{n}} \Big)$.
By analogous reasoning,
$u_{1}\Big( s_{- A},s_{m,k_{m}^{*}},s_{m + 1},$ $_{{\tilde{k}}_{m + 1}},\ldots,s_{n,{\tilde{k}}_{n}} \Big) < u_{1}\Big( s_{- A},s_{m,k_{m}^{*}},s_{{m + 1},k_{m + 1}^{*}},s_{m + 2,{\tilde{k}}_{m + 2}},\ldots,s_{n,{\tilde{k}}_{n}} \Big)$.
By another inductive reasoning and by a chain of inequalities,
$u_{1}\left( s_{- A},{\tilde{s}}_{A} \right) < u_{1}( s_{- A},$ $s_{A}^{*} )$
as we wanted to show. Moreover, if \(k = 1\), \(s_{1,2}\) dominates
\(s_{1,1}\), then
\(u_{1}\left( s_{- A},{\tilde{s}}_{A} \right) < u_{1}\left( 2,s_{- A} \setminus \left\{ s_{1,1} \right\},{\tilde{s}}_{A} \right)\),
and by the joint cooperation of 1 and $m$, we have
\(u_{1}\left( 2,s_{- A} \setminus \left\{ s_{1,1} \right\},{\tilde{s}}_{A} \right) < u_{1}\left( s_{- A},s_{m,{\tilde{k}}_{m} - 1},s_{m + 1,{\tilde{k}}_{m + 1}},\ldots,s_{n,{\tilde{k}}_{n}} \right) \Rightarrow u_{1}\left( s_{- A},{\tilde{s}}_{A} \right) < u_{1}\left( s_{- A},s_{m,{\tilde{k}}_{m} - 1},s_{m + 1,{\tilde{k}}_{m + 1}},\ldots \right.\ \),\(\left. \ s_{n,{\tilde{k}}_{n}} \right)\).
By repeating all the above arguments, we obtain that
\(u_{1}\left( s_{- A},{\tilde{s}}_{A} \right) < u_{1}\left( s_{- A},s_{A}^{*} \right)\).
The test for players \(2,\ldots,\text{ }m - 1\) is analogous. $\square$

\vspace{1\baselineskip}

We also prove the following result: in any prisoner's dilemma, as long as no player has more strategies than all others, the utility each player receives in the cooperative equilibrium is greater than that received in the Nash equilibrium of the game. While intuitive, this result is not only useful for subsequent proofs but is also necessary for our games to be well-defined. Recall that in the Losing Contracts section, when we defined the prisoner's dilemma for three players with two strategies each, we had required this as a condition. We shall now see that this is satisfied by the game's very definition.

\vspace{1\baselineskip}

\noindent \textbf{Lemma 4.} \textit{In an $n$-player prisoner's dilemma in which each player \(i\)
has \(k_{i}\) strategies, if there is no player \(i\) such that
\(k_{i} > k_{j}\) \(\forall j \neq i\) (i.e., the greatest number of
strategies are shared by at least two players), it holds that
\(u_{i}\left( k_{1},\ldots,k_{n} \right) < u_{i}(E)\) \(\forall 1 \leq i \leq n\),
where \(E = \left( s_{1,1},\ldots,s_{n,1} \right)\).}
\smallskip

\noindent \textit{Proof.} Let \(G\) be an $n$-player prisoner's dilemma in which each player \(i\)
has \(k_{i}\) strategies. Without losing generality, suppose
\(k_{1} = k_{2} \geq k_{i}\) \(\forall 3 \leq i \leq n\). Let us consider
induced game \(G_{- \{ 3,\ldots,n\}}\left( k_{3},\ldots,k_{n} \right)\).
For the joint cooperation of 1 and 2,
\(u_{1}\left( k_{1},\ldots,k_{n} \right) < u_{1}\left( k_{1} - 1,k_{2} - 1,k_{3},\ldots,k_{n} \right) < \ldots < u_{1}\left( 1,1,k_{3},\ldots,k_{n} \right)\).
By Lemma 3,
\(u_{1}\left( 1,1,k_{3},\ldots,k_{n} \right) < u_{1}(E) \Rightarrow u_{1}\left( k_{1},\ldots \right.\ \),
\(\left. \ k_{n} \right) < u_{1}(E)\). The case of player 2 is similar
to the previous one. We have \(3 \leq i \leq n\). Now let us consider
induced game
\(G_{- \{ 2,\ldots,n\} \setminus \{ i\}}\left( k_{2},\ldots,k_{i - 1},k_{i + 1},\ldots,k_{n} \right)\).
Due to the joint cooperation of 1 and
\(i\), \(u_{i}\left( k_{1},\ldots,k_{n} \right) < u_{i}( k_{1} - 1,k_{2},\ldots,k_{i - 1},k_{i} - 1,k_{i + 1},\) \(\ldots,k_{n}) < \ldots < u_{i}\left( k_{1} - k_{i} + 1,k_{2},\ldots,k_{i - 1},1,k_{i + 1},\ldots,k_{n} \right)\).
By Lemma 3 again,
\(u_{i}\left( k_{1} - k_{i} + 1,k_{2},\ldots,k_{i - 1},1,k_{i + 1},\ldots,k_{n} \right) < u_{i}(E) \Rightarrow u_{i}\left( k_{1},\ldots,k_{n} \right) < u_{i}(E)\). $\square$

\vspace{1\baselineskip}

We would like no incentives to exist for coalitions before signing the
contract. To this end, for each subset \(A \subseteq \{ 1,\ldots,n\}\)
of players, it must be satisfied that the rest have no joint strategy
\(s_{- A} \in S_{- A}\) such that
\(\sum_{i \notin A}^{}\mspace{2mu} u_{i}\left( s_{- A},\left( k_{j} \right)_{j \in A} \right) > \sum_{i \notin A}^{}\mspace{2mu} u_{i}(1,\ldots,1)\).
Otherwise, we would have a coalition
\(\{ 1,\ldots,n\} \setminus A\) of players who do not even want to
sign the contract, because if they sign another internal contract to
share the utility (remember that it is transferable) among them in an
appropriate way by choosing \(s_{- A}\) jointly, they can all obtain a
higher utility in \(M\) than they would obtain in \(\tilde{M}\) by
signing the original contract (assuming that it optimizes the
cooperative equilibrium); even considering the worst case, where the
rest of the players choose their least cooperative strategies according
to Definition 4. This would generate coalitional incentives. With the
following result, we can see that in the prisoner's dilemma, this is
never the case.

\vspace{1\baselineskip}

\noindent \textbf{Lemma 5.} \textit{If there is a contract prior to an 
-player prisoner's dilemma with finite strategy sets optimizing the cooperative equilibrium, then there are no incentives for coalitions to form before signing the contract, even under transferable utility.}
\smallskip

\noindent \textit{Proof.} First, it is assumed that all players have two strategies each. We
set \(n = 3\) (there is no point in analyzing \(n = 2\) as we assume
that the cooperative equili\-brium is Pareto-optimal). The coalition must
be formed by more than one player (since there is a contract optimizing $E$, so each player obtains a higher utility in the modified game's equilibrium than in the original); and since the cooperative equilibrium
is Pareto-optimal, it also cannot consist of all players (these two properties hold for any
$n$), so we must analyze coalitions of more than one and less than $n$ players. In
this case, that means coalitions of two players. Without loss of
generality, we will analyze what happens if \(A = \{ 3\}\). It si sufficient to show that
\(u_{1}\left( s_{- A},2 \right) + u_{2}\left( s_{- A},2 \right) \leq u_{1}(E) + u_{2}(E)\) (with \(E = (1,1,1))\) \(\forall s_{- A} \in S_{- A} = \{(1,1),(1,2),(2,1),(2,2)\}\). By Lemma 4, we know that $u_{1}(2,2,2)<u_{1}(E)$ and $u_{2}(2,2,2)<u_{2}(E)$, so
\(u_{1}(2,2,2) + u_{2}(2,2,2) < u_{1}(E) + u_{2}(E)\). Moreover, by
\(G_{- \{ 3\}}(2)\), we have \(u_{1}(1,2,2) < u_{1}(2,2,2) < u_{1}(1,1,2) < u_{1}(2,1,2)\),
so we have
\(u_{1}(1,2,2) + u_{2}(1,2,2) < u_{1}(2,1,2) + u_{2}(1,2,2)\). The
analysis of \(s_{- A} = (2,1)\) is analogous to that of
\(s_{- A} = (1,2)\), so it is sufficient to see that
\(u_{1}(2,1,2) + u_{2}(1,2,2) < u_{1}(E) + u_{2}(E)\). As
\(G_{- \{ 2\}}(1)\) is a prisoner's dilemma, for the joint cooperation
between 1 and 3, \(u_{1}(2,1,2) < u_{1}(E)\). Ana\-logously, as
\(G_{- \{ 1\}}(1)\) is also a prisoner's dilemma,
\(u_{2}(1,2,2) < u_{2}(E)\), so the inequa\-lity is strictly enforced.
Now, \(n \geq 3\) is set and we can assume that there are no coalition
incentives for the $n$-player prisoner's dilemmas with two
strategies each. Let \(G\) be a prisoner's dilemma of $n+1$ players with
two strategies each. Let us show that there are no coalitional incentives in \(G\). Without
losing generality, let \(A = \{ m,\ldots,n + 1\}\) be a subset of players
(with \(3 \leq m \leq n + 1\)). \(G_{- \{ n + 1\}}(2)\) has $n$ players, so
given \(s_{- A} \in S_{- A}\), we are left with
\(\sum_{i = 1}^{m - 1}\mspace{2mu} u_{i}\left( s_{- A},2,\ldots,2,2 \right) \leq \sum_{i = 1}^{m - 1}\mspace{2mu} u_{i}(1,\ldots,1,2)\)
(if \(m = n + 1\) we use Lemma 3 and Lemma 4 directly: if \(i \in A\)
chooses
\(s_{i,1} \Rightarrow u_{i}\left( s_{- A},2,\ldots,2 \right) < u_{i}(E)\)
by Lemma 3, and if \(i \in A\) chooses
\(s_{i,2} \Rightarrow u_{i}\left( s_{- A},2,\ldots,2 \right) < u_{i}(E)\)
by Lemma 4, since at least \(n + 1\) chose \(s_{n + 1,2}\)). Given
\(1 \leq i \leq m - 1\), due to the dominance of
\(s_{i,2}\), \(u_{i}(1,\ldots,1,2) < u_{i}\left( 1,\ldots,1,s_{i,2},1,\ldots,1,2 \right)\).
Furthermore, if we just look at the players \(i\) and
\(n + 1,G_{- \{ 1,\ldots,i - 1,i + 1,\ldots,n\}}(1,\ldots,1)\) is a
prisoner's dilemma, so due to the joint cooperation of \(i\) and
\(n + 1\), \(u_{i}\left( 1,\ldots,1,s_{i,2},1,\ldots,1,2 \right) < u_{i}(E)\)
(where from now on \(E = (1,\ldots,1)\)). Because of a chain of
inequalities, we have
$\sum_{i = 1}^{m - 1}\mspace{2mu} u_{i}( s_{- A},2,$ $\ldots,2,2) < \sum_{i = 1}^{m - 1}\mspace{2mu} u_{i}(E)$,
which is what we wanted to prove. By induction on $n$, there are no
coalitional incentives in a prisoner's dilemma in which each player has
two strategies. Let \(G\) be an $n$-player prisoner's dilemma where each player
\(i\) has \(k_{i} \geq 2\) strategies and \(A = \{ m,\ldots,n\}\) be a
subset of players (with \(3 \leq m \leq n\)). Let us define
\(\tilde{k} = \min\left\{ k_{m},\ldots,k_{n} \right\}\). For
each \(2 \leq k \leq \tilde{k}\), considering \(k - 1\) and \(k\) as
the only strategies, we have $k-1$ prisoner's dilemmas of $n$ players in
which each player has two strategies. Given \(s_{- A} \in S_{- A}\), by the result we already proved, we have
\(\sum_{i = 1}^{m - 1}\mspace{2mu} u_{i}\left( s_{- A},\tilde{k},\ldots,\tilde{k} \right) < \sum_{i = 1}^{m - 1}\mspace{2mu} u_{i}(\tilde{k} - 1,\ldots,\tilde{k} - 1)\),
but
(\(s_{1,\tilde{k} - 1},\ldots\),\(\left. \ s_{m - 1,\tilde{k} - 1} \right) \in S_{- A} \Rightarrow\)
by inductive reasoning,
\(\sum_{i = 1}^{m - 1}\mspace{2mu} u_{i}(\tilde{k} - 1,\ldots,\tilde{k} - 1) < \sum_{i = 1}^{m - 1}\mspace{2mu} u_{i}(\tilde{k} - 2,\ldots,\tilde{k} - 2) < \ldots < \sum_{i = 1}^{m - 1}\mspace{2mu} u_{i}(E)\).
Through a chain of inequalities, we get
\(\sum_{i = 1}^{m - 1}\mspace{2mu} u_{i}\left( s_{- A},\tilde{k},\ldots,\tilde{k} \right) < \sum_{i = 1}^{m - 1}\mspace{2mu} u_{i}(E)\).
If \(k_{i} = \tilde{k}\) \(\forall m \leq i \leq n\), we are done. If
this is not the case, by Lemma 3, for each
\(1 \leq i \leq m - 1\), \(u_{i}\left( s_{- A},k_{m},\ldots,k_{n} \right) < u_{i}\left( s_{- A},\tilde{k},\ldots,\tilde{k} \right)\), so we have
\(\sum_{i = 1}^{m - 1}\mspace{2mu} u_{i}\left( s_{- A},k_{m},\ldots,k_{n} \right) < \sum_{i = 1}^{m - 1}\mspace{2mu} u_{i}\left( s_{- A},\tilde{k},\ldots,\tilde{k} \right) < \sum_{i = 1}^{m - 1}\mspace{2mu} u_{i}(E)\) with a strict inequality. $\square$

\vspace{1\baselineskip}

By Lemma 2 and Lemma 5, we already have that our game is well defined
and that in the latter, it is in the players' interest to sign a
contract that optimizes the cooperative equilibrium. Our goal will be to
define a contract that makes this equilibrium a strong Nash equilibrium. Because of how we define the prisoner's dilemma, we would also
want the contract in question to maintain the incentives to cooperate in
each restricted game.

\vspace{1\baselineskip}

\noindent \textbf{Definition 6.} Let \(G\) be a game where \(S\) is the set of joint strategies
of all players. We will say that \(\tilde{G}\) is a restricted game
of \(G\) if \(\tilde{S} \subseteq S\) is the set of joint strategies
of all players in \(\tilde{G}\), where each set of unrestricted
strategies \({\tilde{S}}_{i}\) of each player \(i\) always satisfies
that its strategies are contiguous two by two (i.e.,
\(s_{i,k - 1},s_{i,k + 1} \in {\tilde{S}}_{i} \Rightarrow s_{i,k} \in {\tilde{S}}_{i}\) \(\forall k\)).
In particular, induced games are restricted games. The restricted game
given by \(s_{A} \in S_{A}\) of a subset of players \(A\), and a
restricted set of joint strategies \({\tilde{S}}_{- A} \in S_{- A}\)
from \(- A\), is denoted by \(G_{- A}\left( s_{A},{\tilde{S}}_{- A} \right)\).
Beyond the general definition, we will consider in our games that if
\(s_{i,k + 1}\) is a restricted strategy, then \(s_{i,k}\) is also a
restricted strategy.

\vspace{1\baselineskip}

\noindent \textbf{Lemma 6.} \textit{In an $n$-player prisoner's dilemma in which each player \(i\)
has \(k_{i}\) strategies, every restricted game is a prisoner's dilemma.}
\smallskip

\noindent \textit{Proof.} Without loss of generality, let \(A \subseteq \{ 1,\ldots,n\}\) be such
that
\(1 \leq |A| \leq n - 2\) and \(s_{A} \in S_{A},{\tilde{S}}_{- A} \subseteq S_{- A}\).
By Definition 5, \(G_{- A}\left( s_{A} \right)\) is a prisoner's
dilemma, which means that every two-player induced game from \(G_{- A}\left( s_{A} \right)\) is a priso\-ner's dilemma. In particular,
by restricting the strategies of all these induced two-player games to
\({\tilde{S}}_{- A}\), by Definition 6, they remain
prisoner's dilemmas. Therefore, every two-player induced game in
restricted game \(G_{- A}\left( s_{A},{\tilde{S}}_{- A} \right)\) is
a prisoner's dilemma; and by definition, so is this last one. $\square$

\vspace{1\baselineskip}

We already proved that in the prisoner's dilemma, there are no coalitional incentives prior to signing a contract that optimizes the cooperative equilibrium. As
mentioned before, our goal is to prevent such incentives after the
contract is signed: neither in the game itself, nor in any of the restricted games.

\vspace{1\baselineskip}

\noindent \textbf{Theorem 1.} \textit{Let \(G\) be an $n$-player prisoner's dilemma in which each player
\(i\) has \(k_{i}\) strategies, and let \(C\) be a losing contract such that
$r_{i,k} = \sum_{m = 2}^{k}\mspace{2mu}\max_{{\tilde{s}}_{- i} \in S_{- i}}$ \(\{ u_{i}\left( m,{\tilde{s}}_{- i} \right)- u_{i}\left( m - 1,{\tilde{s}}_{- i} \right) \} + \varepsilon_{i,k}\) (with \(\varepsilon_{i,k} > \varepsilon_{i,k - 1} > \ldots > \varepsilon_{i,2} > 0)\)
for each \(1 \leq i \leq n,2 \leq k \leq k_{i}\). Then, each \(r_{i,k}\)
is in its feasible region, \(C\) optimizes
\(E = \left( s_{1,1},\ldots,s_{n,1} \right)\), and this last equilibrium
is unique. Moreover, it is a strong equilibrium. Furthermore, \(C\) makes the joint cooperation in
each restricted game of \(G\) a unique and strong Nash equilibrium.}
\smallskip

\noindent \textit{Proof.} Recall that in Lemma 1, we defined ${\tilde{r}}_{i,k} = \max_{{\tilde{s}}_{- i}} \in S_{- i}\{ u_{i}\left( k,{\tilde{s}}_{- i} \right) -$ $u_{i}\left( 1,{\tilde{s}}_{- i} \right) \} + {\tilde{\varepsilon}}_{i,k}$
for each \(1 \leq i \leq n,2 \leq k \leq k_{i}\) and that the result
holds for any \({\tilde{\varepsilon}}_{i,k} > 0\). Let's compare
these values with our new \(r_{i,k}\). Taking into account Definition 4,
for each \(1 \leq i \leq n,2 \leq k \leq k_{i}\), strategy \(s_{i,k}\)
dominates strategy \(s_{i,k - 1}\). Fix 
\(1 \leq i \leq n\) and \(2 \leq k \leq k_{i}\).
Let \(s_{- i} \in S_{- i}\) be such that
\({\max}_{{\tilde{s}}_{- i} \in S_{- i}}\left\{ u_{i}\left( k,{\tilde{s}}_{- i} \right) - u_{i}\left( 1,{\tilde{s}}_{- i} \right) \right\} = u_{i}\left( k,s_{- i} \right) - u_{i}(1\),
\(s_{- i}\)). We have
\(r_{i,k} > \sum_{m = 2}^{k}\mspace{2mu}{\max}_{{\tilde{s}}_{- i} \in S_{- i}}\left\{ u_{i}\left( m,{\tilde{s}}_{- i} \right) - u_{i}\left( m - 1,{\tilde{s}}_{- i} \right) \right\}\)
(because \(\varepsilon_{i,k} > 0) \geq \sum_{m = 2}^{k}\mspace{2mu}\left\lbrack u_{i}\left( m,s_{- i} \right) - u_{i}\left( m - 1,s_{- i} \right) \right\rbrack = u_{i}\left( k,s_{- i} \right) - u_{i}\left( 1,s_{- i} \right)\)
(it is a finite telescopic summation). Then, taking
\({\tilde{\varepsilon}}_{i,k} = r_{i,k} - \left\lbrack u_{i}\left( k,s_{- i} \right) \right.\ \left. \  - u_{i}\left( 1,s_{- i} \right) \right\rbrack > 0\),
we have the conditions of Lemma 1 (since our \(r_{i,k}\) is equivalent
to \({\tilde{r}}_{i,k}\) for \({\tilde{\varepsilon}}_{i,k} > 0\)
and \(E\) is a cooperative equilibrium of \(G\)); therefore, \(r_{i,k}\)
is in its feasible region and \(C\) optimizes \(E\) and makes the latter
the unique Nash equilibrium. Now we can prove the following: after
signing \(C\), given \(1 \leq i \leq n,2 \leq k \leq k_{i}\), it is
satisfied that \(s_{i,k - 1}\) dominates \(s_{i,k}\). For this, it is
enough to see that given any \(s_{- i} \in S_{- i}\), then
\(u_{i}\left( k,s_{- i} \right) - r_{i,k} < u_{i}\left( k - 1,s_{- i} \right) - r_{i,k - 1}\).
This is fulfilled
\(\Leftrightarrow u_{i}\left( k,s_{- i} \right) - \sum_{m = 2}^{k}\mspace{2mu}\left\lbrack {\max}_{{\tilde{s}}_{- i} \in S_{- i}}\left\{ u_{i}\left( m,{\tilde{s}}_{- i} \right) - u_{i}\left( m - 1,{\tilde{s}}_{- i} \right) \right\} \right\rbrack - \varepsilon_{i,k} < u_{i}(k - 1\),\(\left. \ s_{- i} \right) - \sum_{m = 2}^{k - 1}\mspace{2mu}\left\lbrack {\max}_{{\tilde{s}}_{- i} \in S_{- i}}\left\{ u_{i}\left( m,{\tilde{s}}_{- i} \right) - u_{i}\left( m - 1,{\tilde{s}}_{- i} \right) \right\} \right\rbrack - \varepsilon_{i,k - 1}\)
(where the last summation and \(\varepsilon_{i,k - 1}\) are zero if \(k = 2\), because $r_{i,1} = \varepsilon_{i,1} = 0 ) \Leftrightarrow u_{i}\left( k,s_{- i} \right) - \lbrack
\max_{{\tilde{s}}_{- i} \in S_{- i}}\left\{ u_{i}\left( k,s_{- i} \right) - u_{i}\left( k - 1,s_{- i} \right) \right\} - \varepsilon_{i,k} < u_{i}\left( k - 1,s_{- i} \right) - \varepsilon_{i,k - 1}$,
but this is
valid because
\(\varepsilon_{i,k} > \varepsilon_{i,k - 1}\left( \varepsilon_{i,2} > 0 \right.\ \)
if \(\left. \ k = 2 \right)\) and
$u_{i}\left( k,s_{- i} \right) - u_{i}( k - 1,$ $s_{- i}) \leq\max_{{\tilde{s}}_{- i} \in S_{- i}}\left\{ u_{i}\left( k,s_{- i} \right) - u_{i}\left( k - 1,s_{- i} \right) \right\}$,
so after signing \(C,s_{i,k - 1}\) dominates
\(s_{i,k}\) \(\forall 1 \leq i \leq n,2 \leq k \leq k_{i}\) (this is not sufficient: recall the modified game presented in Tables~\ref{tab:matrix_7} and~\ref{tab:matrix_8}). Without loss of
generality, let us consider the restricted game
\(G_{- A}\left( s_{A},{\tilde{S}}_{- A} \right)\), where
\(A = \{ m + 1,\ldots,n\}\), \(s_{A} \in S_{A}\) and \({\tilde{S}}_{- A} \subseteq S_{- A}\)
(with \(2 \leq m \leq n - 1)\), and where for each \(1 \leq i \leq m\), its
possible strategies within \({\tilde{S}}_{- A}\) are
\(s_{i,k_{i,1}},\ldots,s_{i,k_{i}}\) with
\(k_{i,1} < \ldots < k_{i}\left( 1 \leq k_{i,1} < k_{i} \right)\). Based
on what we already proved, \(s_{i,k_{i,1}}\) is dominant for \(i\) at
\(G_{- A}\left( s_{A},{\tilde{S}}_{- A} \right)\) for each
\(1 \leq i \leq m\); thus,
\(\left( s_{1,k_{1,1}},\ldots,s_{m,k_{m,1}} \right)\) is the only Nash
equilibrium of this restricted game. Moreover, without loss of
generality, we define
\(B = \{ 1,\ldots,\tilde{m}\}\) and fix \(s_{B} \in {\tilde{S}}_{B}\) (with \(2 \leq \tilde{m} \leq m\),
where \({\tilde{S}}_{B}\) is the set of joint strategies of \(B\)
restricted to \({\tilde{S}}_{- A}\)). Suppose that at least two
\(i \in B\) do not choose \(s_{i,k_{i,1}}\) at \(s_{B}\) (if only one
does not choose \(s_{i,k_{i,1}}\), as \(s_{i,k - 1}\) dominates
\(s_{i,k}\) after signing \(C\), and by Lemma 3, all players of
\(B\) get less utility than if no one deviates). Before signing \(C\),
restricted game
\(G_{- \{\tilde{m} + 1,\ldots,n\}}\left( k_{\tilde{m} + 1,1},\ldots,k_{m,1},s_{A},S_{B}^{*} \right)\)
was a prisoner's dilemma under Lemma 6 (in
\(S_{B}^{*} \subseteq {\tilde{S}}_{B}\) the least cooperative
strategies are the ones chosen at \(s_{B}\), and the most cooperative
are still the ones \(s_{i,k_{i,1}}\)). Assuming without loss of
gene\-rality that each player \(i \in B \setminus \{\tilde{m}\}\)
has more (or at least the same) number of strategies at \(S_{B}^{*}\)
than player \(i + 1\), by Lemma 4,
\(u_{i}\left( s_{B},k_{\tilde{m} + 1,1},\ldots,k_{m,1},s_{A} \right) < u_{i}\left( k_{1}^{*} - \left( k_{2}^{*} - k_{2,1} \right),k_{2,1},\ldots,k_{\tilde{m},1},k_{\tilde{m}+1,1},\ldots,k_{m,1},s_{A} \right)\)
for each \(1 \leq i \leq \tilde{m}\), where
\(s_{1,k_{1}^{*}},s_{2,k_{2}^{*}}\) are the strategies chosen by 1 and 2 at
\(s_{B}\). Under Lemma 3, for each
\(2 \leq i \leq \tilde{m}\), \(u_{i}\left( k_{1}^{*} - \left( k_{2}^{*} - k_{2,1} \right),k_{2,1}\ldots,k_{\tilde{m},1},k_{\tilde{m} + 1,1},\ldots,k_{m,1},s_{A} \right) \leq u_{i}\left( k_{1,1},\ldots,k_{m,1},s_{A} \right)\),
since \(k_{1}^{*} - k_{1,1} \geq k_{2}^{*} - k_{2,1}\) (player 1 has
equal or more strategies than player 2 at \(S_{B}^{*}\)) $\Rightarrow$
\(k_{1}^{*} - \left( k_{2}^{*} - k_{2,1} \right) \geq k_{1,1},\) so
$u_{i}( s_{B},k_{\tilde{m} + 1,1},\ldots,$ $k_{m,1},s_{A} ) < u_{i}\left( k_{1,1},\ldots,k_{m,1},s_{A} \right)$ $\forall 2 \leq i \leq \tilde{m}$.
Noting that \(r_{i,k} > r_{i,\tilde{k}}\) if \(k > \tilde{k}\)
by definition, we infer that the last inequality holds after signing
\(C\), as does that of player 1. In addition, we already saw that
\(s_{1,k - 1}\) dominates \(s_{1,k}\) after signing \(C\), so
\(u_{1}\left( k_{1}^{*} - \left( k_{2}^{*} - k_{2,1} \right),k_{2,1}\ldots,k_{\tilde{m},1},k_{\tilde{m} + 1,1},\ldots,k_{m,1},s_{A} \right) \leq u_{1}\left( k_{1,1},\ldots,k_{m,1},s_{A} \right) \Rightarrow u_{1}\left( s_{B},k_{\tilde{m} + 1,1},\ldots,k_{m,1},s_{A} \right) < u_{1}\left( k_{1,1},\ldots,k_{m,1},s_{A} \right)\)
after signing \(C\). Then we have
$\sum_{i = 1}^{\tilde{m}}\mspace{2mu} u_{i}\left( s_{B},k_{\tilde{m} + 1,1},\ldots,k_{m,1},s_{A} \right) < \sum_{i = 1}^{\tilde{m}}\mspace{2mu} u_{i}$ $\left( k_{1,1},\ldots,k_{\tilde{m},1},k_{\tilde{m} + 1,1},\ldots,k_{m,1},s_{A} \right)$.
This proves that (\(s_{1,k_{1,1}},\ldots,s_{m,k_{m,1}}\)) is a strong Nash equilibrium in \(G_{- A}\left( s_{A},{\tilde{S}}_{- A}\right)\) even with transferable utility, since
\(B \subseteq - A,s_{B} \in {\tilde{S}}_{B}\) were arbitrary. Since
\(G_{- A}\left( s_{A},{\tilde{S}}_{- A} \right)\) was also an arbitrary
restricted game of \(G\), \(C\) makes the joint cooperation in
each restricted game a unique and strong Nash equilibrium.
Finally, by analogous reasoning, \(E\) is a strong Nash equilibrium in $G$, since all the analysis we did in \(G_{- A}\left( s_{A},{\tilde{S}}_{- A}\right)\) still holds in \(G\) without considering any
restriction. This can be seen by defining \(B = \{ 1,\ldots,\tilde{m}\}\) with
\(2 \leq \tilde{m} \leq n\) and taking any \(s_{B} \in S_{B}\). $\square$

\vspace{1\baselineskip}

\noindent \textbf{Remark 3.} In the above result, as in Lemma 5, we assume that
$E = ( s_{1,1},\ldots,$ $s_{n,1} )$ is Pareto-optimal under
transferable utility; i.e., that
$\sum_{i = 1}^{n}\mspace{2mu} u_{i}(P)\leq \sum_{i = 1}^{n}$ $u_{i}(E)$
for all strategy profiles \(P \neq E\), although actually we need not
ask for this condition: note that if each player \(i\) chooses the
strategy \(s_{i,{\tilde{k}}_{i}}\) at \(P\), with
\({\tilde{k}}_{i} \geq 2\) for at least two
\(i \in \{ 1,\ldots,n\}\), then (assuming without loss of generality
that
\({\tilde{k}}_{i} \geq {\tilde{k}}_{i + 1}\) \(\forall 1 \leq i \leq n - 1\))
we have under Lemma 4 that
\(u_{i}(P) < u_{i}\left( {\tilde{k}}_{1} - {\tilde{k}}_{2} + 1,1,\ldots,1 \right)\forall 1 \leq i \leq n\).
Specifically, if
\({\tilde{k}}_{1} = {\tilde{k}}_{2} \Rightarrow \sum_{i = 1}^{n}\mspace{2mu} u_{i}(P) < \sum_{i = 1}^{n}\mspace{2mu} u_{i}(E)\).
These two inequalities show that for \(E\) to be a Pareto-optimal
profile, it is enough to require
\(\sum_{i = 1}^{n}\mspace{2mu} u_{i}(P) < \sum_{i = 1}^{n}\mspace{2mu} u_{i}(E)\)
for any profile \(P\) such that a single player deviates from \(E\); if
the latter is satisfied due to the first of the two previous
inequalities, we have
\(\sum_{i = 1}^{n}\mspace{2mu} u_{i}(P) < \sum_{i = 1}^{n}\mspace{2mu} u_{i}(E)\)
for any profile \(P \neq E\). Therefore, it is enough to require a much
weaker condition for \(E\) to be Pareto-optimal in prisoner's dilemmas.

\vspace{1\baselineskip}

Recall that our goal in seeking a contract that achieves a strong equilibrium in the joint cooperation of each restricted game was to account for potential external restrictions to the game that could decrease the number of strategies available to any subset of players. By Lemma 6, this possible restriction still leaves all players with more than one strategy in yet another new prisoner's dilemma. We have already seen in Lemma 4 that, in such a case, joint cooperation yields higher payoffs to the players than when they all defect (except when one player has a strictly larger number of strategies than all the others). 

We have yet to solve the problem of ensuring that signing the contract remains a dominant strategy, despite the players' uncertainty regarding whether their most cooperative strategies will remain available due to potential restrictions. Before continuing, let us prove the following result and then define the \textit{reduced amounts}.

\vspace{1\baselineskip}

\noindent \textbf{Lemma 7.} \textit{In an $n$-player prisoner's dilemma in which each player \(i\)
has \(k_{i}\) strategies, given \(1 \leq i \leq n,1 \leq k < k_{i}\), it
holds that
\(u_{i}\left( s^{*} \right) < u_{i}\left( k,s_{- i} \right)\) \(\forall s_{- i} \in S_{- i} \setminus S_{- i,k}\),
where \(S_{- i,k} := \{ s_{- i} \in S_{- i}\): every
\(j \neq i\) chooses \(s_{j,\tilde{k}}\) at \(s_{- i}\) for some
\(\tilde{k}\) such that
\(k_{j} - \tilde{k} < k_{i} - k\}\), and
\(s^{*} := \left( s_{1,k_{1}},\ldots,s_{n,k_{n}} \right)\).}
\smallskip

\noindent \textit{Proof.} Let \(1 \leq i \leq n,1 \leq k < k_{i},s_{- i} \in S_{- i} \setminus S_{- i,k}\).
Note that \(\exists j \neq i\) such that \(j\) chooses
\(s_{j,\tilde{k}}\) at \(s_{- i}\) for some \(\tilde{k}\) such
that \(k_{j} - \tilde{k} \geq k_{i} - k\), so
\(u_{i}\left( s^{*} \right) < u_{i}\left( k,{\tilde{s}}_{- i} \right)\),
where at \({\tilde{s}}_{- i}\), all players choose their
least cooperative strategy except player \(j\), who chooses
\(s_{j,k_{j} - k_{i} + k}\). Since
\(k_{j} - k_{i} + k \geq \tilde{k}\), by Lemma 3 we have that
\(u_{i}\left( k,{\tilde{s}}_{- i} \right) \leq u_{i}\left( k,s_{- i} \right) \Rightarrow u_{i}\left( s^{*} \right) < u_{i}\left( k,s_{- i} \right)\). $\square$

\vspace{1\baselineskip}

\noindent \textbf{Definition 7.} Let \(G\) be an $n$-player prisoner's dilemma in which each
player \(i\) has \(k_{i}\) strategies, \(r_{i,k} = c\) be the amount for player $i$ when choosing the strategy \(k\) in a losing contract
\(\left( 1 \leq i \leq n,2 \leq k \leq k_{i} \right),s^{*} = \left( s_{1,k_{1}},\ldots,s_{n,k_{n}} \right)\).
We will say that \(r_{i,k}^{*}\) is the reduced amount of player \(i\)
when choosing the strategy \(k\) if: 
\begin{itemize}
    \item[$\bullet$] is equal to \(c\) whenever
\(u_{i}\left( s^{*} \right) < u_{i}\left( k,s_{- i} \right) - c\),
    \item[$\bullet$] it is the amount necessary for
\(u_{i}\left( s^{*} \right) = u_{i}\left( k,s_{- i} \right) - r_{i,k}^{*}\) whenever
\(u_{i}\left( k,s_{- i} \right)\) \(- c \leq u_{i}\left( s^{*} \right) < u_{i}\left( k,s_{- i} \right)\)
(i.e., in this case,
\(\left. \ r_{i,k}^{*} = u_{i}\left( k,s_{- i} \right) - u_{i}\left( s^{*} \right) \right)\),
    \item[$\bullet$] it is zero whenever
\(u_{i}\left( k,s_{- i} \right) \leq u_{i}\left( s^{*} \right)\).
\end{itemize}
Note that the value of \(r_{i,k}^{*}\) varies depending on \(s_{- i}\) in the
second case. Henceforth, writing \(r_{i,k}^{*} = c\) means that it represents the reduced amount satisfying the above conditions.

It can be seen that with these reduced amounts, the utilities in the new game can be explicitly written in terms of their three possible cases.

\vspace{1\baselineskip}

\noindent \textbf{Definition 8.} Let \(r_{i,k}^{*}\) be a reduced amount. We define the utility of player \(i\) in choosing strategy \(k\) after the losing
contract as
\(u_{i}^{*}\left( k,s_{- i} \right) = u_{i}\left( k,s_{- i} \right) - r_{i,k}^{*}\).
Then, by how \(r_{i,k}^{*}\) is defined in each case, we have that
$u_{i}^{*}\left( k,s_{- i} \right) = \max\{ u_{i}\left( k,s_{- i} \right) - r_{i,k},$ $\min\left\{ u_{i}\left( s^{*} \right),u_{i}\left( k,s_{- i} \right) \right\}\}$,
where \(r_{i,k} = c\) in the first value and
\(u_{i}\left( s^{*} \right)\) in the second are fixed, and
\(u_{i}\left( k,s_{- i} \right)\) (in the first and third values) is the
value of the pre-contract utility that we already know.

\vspace{1\baselineskip}

While the reduced amounts implicitly depend on the strategies of others, the payoffs in the new game can be expressed without taking others' strategies into account, provided the pre-contract payoffs of the game are known. The amounts are decreased to ensure that no payoff falls below the equilibrium of the original game, except for those that were already below it prior to the contract. This reduction in the corresponding payoffs ensures that signing the contract is a dominant strategy.

A challenge arises because we require \(u_{i}\left( s^{*} \right) < u_{i}\left( 1,s_{- i} \right)\) \(\forall s_{- i} \in S_{- i} \setminus \left\{ s_{- i}^{*} \right\}\)
(where
\(s_{- i}^{*} = \left( s_{1,k_{1}},\ldots,s_{i - 1,k_{i - 1}},s_{i + 1,k_{i + 1}},\ldots,s_{n,k_{n}} \right)\)),
but this condition is not generally satisfied. To address this, we establish the following rule: if a player \(i\) chooses strategy \(s_{i,k}\) and every
\(j \neq i\) chooses \(s_{j,{\tilde{k}}_{j}}\) with
\({\tilde{k}}_{j}\) such that
\(k_{j} - {\tilde{k}}_{j} < k_{i} - k\), then player $i$ is automatically assigned the strategy \(s_{i,k^{*}}\), where
\(k^{*}: = k_{i} - {\max}_{j \neq i}\left\{ k_{j} - {\tilde{k}}_{j} \right\}\).
This not only facilitates the use of Lemma 7 but also provides a payoff guarantee for each player, in the sense that they will be protected when acting as the most cooperative player if no other player matches that same level of cooperation. 

\vspace{1\baselineskip}

\noindent \textbf{Theorem 2.} \textit{Let \(G\) be an $n$-player prisoner's dilemma in which each player
\(i\) has \(k_{i}\) strategies, and let \(C\) be a losing contract such that
$r_{i,k}^{*} = \sum_{m = 2}^{k}\mspace{2mu}\max_{{\tilde{s}}_{- i} \in S_{- i}}$ $\{ u_{i}\left( m,{\tilde{s}}_{- i} \right)-$ $u_{i}\left( m - 1,{\tilde{s}}_{- i} \right)\} + \varepsilon_{i,k}$ (with $\varepsilon_{i,k} > \varepsilon_{i,k - 1} > \ldots > \varepsilon_{i,2} > 0)$
for each \(1 \leq i \leq n,2 \leq k \leq k_{i}\). Then, each
\(r_{i,k}^{*}\) is in its feasible region and \(C\) optimizes
\(E = \left( s_{1,1},\ldots,s_{n,1} \right)\), which is a strong Nash equilibrium. Furthermore, \(C\) ensures
that in every restricted game containing at least one
\(r_{i,k}^{*} = r_{i,k}\) when all other players choose to cooperate, joint cooperation is a strong Nash equilibrium, \smallskip provided that the player who cooperates the most is compensated (excluding ties).}

\noindent \textit{Proof.} We saw in Theorem 1 that, given any
\(1 \leq i \leq n,2 \leq k \leq k_{i}\), if
\(r_{i,k}^{*} = r_{i,k}\Rightarrow\) \(r_{i,k}^{*}\) is in its feasible
region. Furthermore, if
\(r_{i,k}^{*} = u_{i}\left( k,s_{- i} \right) - u_{i}\left( s^{*} \right)\), then
we have
\(u_{i}^{*}\left( k,s_{- i} \right) = u_{i}\left( s^{*} \right) < u_{i}\left( 1,s_{- i} \right)\)
(considering by Lemma 7 the
compensation for \(i\) if applicable), because if
\(s_{- i} = s_{- i}^{*}\), as \(s_{i,k_{i}}\) is dominant for \(i\)
before signing \(C\), we would have
\(u_{i}\left( k,s_{- i} \right) \leq u_{i}\left( s^{*} \right)\), but by definition of \(r_{i,k}^{*}\) we are not in this case. Finally, if \(r_{i,k}^{*} = 0\), the amount does not
affect the utility of player \(i\), so it should not satisfy any conditions.
Therefore, \(r_{i,k}^{*}\) remains in its feasible region regardless of the
values it takes. Moreover, by Lemma 7 (always considering
compensation if applicable), we have that
\(u_{i}\left( s^{*} \right) < u_{i}(k,1\ldots,1) \Rightarrow u_{i}^{*}(k,1,\ldots,1) = u_{i}(k,1,\ldots,1) - r_{i,k}^{*}\) with \(r_{i,k}^{*} \neq 0\).
If
\(r_{i,k}^{*} = u_{i}(k,1,\ldots,1) - u_{i}\left( s^{*} \right) \Rightarrow u_{i}^{*}(k,1,\ldots\),\(1) = u_{i}\left( s^{*} \right) < u_{i}(E)\)
due to Lemma 4 (because owing to the compensation, we can now use it
regardless of the number of strategies). If
\(r_{i,k}^{*} = r_{i,k} \Rightarrow u_{i}^{*}(k,1,\ldots,1) = u_{i}(k,1,\ldots,1) - r_{i,k} < u_{i}(E)\) as we saw in Theorem 1. We have that \(E\) is
a Nash equilibrium, and since there are no amounts for \(k = 1\), then
\(C\) optimizes \(E\). Let us see that after signing \(C\),
\(s_{i,k - 1}\) dominates \(s_{i,k}\) non-strictly in
\(S_{- i} \setminus \left\{ s_{- i}^{*} \right\}\). Let
\(s_{- i} \neq s_{- i}^{*}\). If \(r_{i,k}^{*} = r_{i,k}\) we already saw
in Theorem 1 that it is true. If
\(r_{i,k}^{*} = u_{i}\left( k,s_{- i} \right) - u_{i}\left( s^{*} \right)\),
we have two options: if
\(r_{i,k - 1}^{*} = u_{i}\left( k - 1,s_{- i} \right) - u_{i}\left( s^{*} \right) \Rightarrow u_{i}^{*}\left( k,s_{- i} \right) = u_{i}\left( s^{*} \right) = u_{i}^{*}\left( k - 1,s_{- i} \right)\),
and if
\(r_{i,k - 1}^{*} = r_{i,k - 1} \Rightarrow u_{i}^{*}\left( k - 1,s_{- i} \right) = u_{i}(k\left. \  - 1,s_{- i} \right) - r_{i,k - 1} > u_{i}\left( s^{*} \right) = u_{i}^{*}\left( k,s_{- i} \right)\)
by definition of \(r_{i,k - 1}^{*}\). It cannot be the case that
\(r_{i,k - 1}^{*} = 0\), because we would have to
\(u_{i}\left( k - 1,s_{- i} \right) \leq u_{i}\left( s^{*} \right)\),
which contradicts Lemma 7. Note that \(r_{i,k}^{*}\) cannot be zero neither
by Lemma 7; and if \(k = k_{i}\), by Lemma 3. For the two values
that \(r_{i,k}^{*}\) can take, we had that
\(u_{i}^{*}\left( k,s_{- i} \right) \leq u_{i}^{*}\left( k - 1,s_{- i} \right)\)
as we wanted. Moreover, let us note that
\(u_{i}^{*}\left( k,{\tilde{s}}_{- i} \right) \leq u_{i}^{*}\left( k,s_{- i} \right)\)
if a non-empty subset of players chooses less cooperative strategies
at \({\tilde{s}}_{- i}\) than at \(s_{- i}\): if
\(r_{i,k}^{*} = r_{i,k} \vee r_{i,k}^{*} = 0\) is satisfied under
Lemma 3, and if
\(u_{i}^{*}\left( k,{\tilde{s}}_{- i} \right) = u_{i}\left( s^{*} \right) \Rightarrow u_{i}\left( s^{*} \right) < u_{i}\left( k,{\tilde{s}}_{- i} \right) < u_{i}\left( k,s_{- i} \right) \Rightarrow u_{i}^{*}\left( k,{\tilde{s}}_{- i} \right) = u_{i}\left( s^{*} \right) \leq u_{i}^{*}\left( k,s_{- i} \right)\)
by definition of \(r_{i,k}^{*}\). Let us also note that if \(k = 1\), then
inequality is strict. Without loss of generality, let
\(1 \leq \tilde{m} \leq m,2 \leq m \leq n - 1,A = \{ m + 1,\ldots,n\}\),\(s_{A} \in S_{A},{\tilde{S}}_{- A} \subseteq S_{- A}\),
where the most cooperative strategy of each player \(i\) in $\tilde{S}_{-A}$ is
\(s_{i,k_{i,1}}\) and the least coope\-rative one is \(s_{i,k_{i}}\). Let us
also assume that
\(r_{i,k_{i,1}}^{*} = r_{i,k_{i,1}}\) \(\forall 1 \leq i \leq \tilde{m}\)
and that if
\(\tilde{m} < m,u_{i}^{*}\left( k_{i,1},s_{- i} \right) = u_{i}\left( s^{*} \right)\forall\tilde{m} + 1 \leq i \leq m\),
provided that the other players also choose their most cooperative
strategies at \({\tilde{S}}_{- A}\). Let us consider restricted
game \(G_{- A}\left( s_{A},{\tilde{S}}_{- A} \right)\). Suppose
that a group of players (it can also be one or all of them) deviates
from \(\left( s_{1,k_{1,1}},\ldots,s_{m,k_{m,1}} \right)\) and that
\(s_{A} \neq \left( s_{m + 1,k_{m + 1}},\ldots,s_{n,k_{n}} \right)\)
(we will prove this particular case later). If an arbitrary player \(i\) is from
the deviating group (chooses \(s_{i,k}\) for some \(k > k_{i,1}\)), due
to the two inequalities we obtained, considering that the group
deviates to the joint strategy \(\tilde{s}\), we have that
\(u_{i}^{*}\left( k,{\tilde{s}}_{- i} \right) = u_{i}\left( k,{\tilde{s}}_{- i} \right) - r_{i,k}^{*} \leq u_{i}\left( k,s_{- i} \right) - r_{i,k}^{*} \leq u_{i}\left( k_{i,1},s_{- i} \right) - r_{i,k_{i,1}} = u_{i}^{*}\left( k_{i,1},s_{- i} \right)\), where the latter inequality is strict if
\(1 \leq i \leq \tilde{m}\) (and the former is an equality if it is a one-player group, since we would have
\({\tilde{s}}_{- i} = s_{- i}\)). Moreover, if player \(i\) is from
the non-deviating group, we have
\(u_{i}^{*}\left( k_{i,1},{\tilde{s}}_{- i} \right) \leq u_{i}^{*}\left( k_{i,1},s_{- i} \right)\),
where the inequality is strict if \(1 \leq i \leq \tilde{m}\),
because in this case,
\(u_{i}^{*}\left( k_{i,1},s_{- i} \right) = u_{i}\left( k_{i,1},s_{- i} \right) - r_{i,k} > u_{i}\left( k_{i,1},{\tilde{s}}_{- i} \right) - r_{i,k}^{*} = u_{i}^{*}\left( k_{i,1},{\tilde{s}}_{- i} \right)\)
for any value of \(r_{i,k}^{*}\). We conclude that if
\(B \subseteq - A\), \(\sum_{i \in B}^{}\mspace{2mu} u_{i}^{*}\left( P,E_{- B} \right) < \sum_{i \in B}^{}\mspace{2mu} u_{i}^{*}\left( E_{- A} \right)\)
for all strategy profiles \(P \subseteq {\tilde{S}}_{B}\) (we note
by \({\tilde{S}}_{B}\) to joint strategies restricted to
\({\tilde{S}}_{- A}\)), where \(E_{D}\) is by definition the most cooperative
set of strategies of any set \(D\) taking into account its
restrictions. We just need to see the case in which
\(s_{A} = \left( s_{m + 1,k_{m + 1}},\ldots,s_{n,k_{n}} \right)\). If
the deviating coalition consists of fewer than \(m - 1\) players, the previous
result remains valid, because
\({\tilde{s}}_{- i} \neq s_{- i}^{*}\) \(\forall 1 \leq i \leq m \Rightarrow s_{- i} \neq s_{- i}^{*}\) \(\forall 1 \leq i \leq m\)
(because in \(s_{- i}\) there are more cooperators than in
\({\tilde{s}}_{- i}\)), so it is still valid that
\(u_{i}^{*}\left( k,s_{- i} \right) \leq u_{i}^{*}\left( k - 1,s_{- i} \right)\).
If all players deviate (assuming they deviate to their least cooperative strategy; otherwise, it reduces to the previous case), given
\(1 \leq i \leq m\), \(u_{i}\left( k_{i},{\tilde{s}}_{- i} \right) = u_{i}\left( s^{*} \right) \Rightarrow u_{i}^{*}\left( k_{i},{\tilde{s}}_{- i} \right) = u_{i}\left( s^{*} \right) < u_{i}\left( k_{i,1},s_{- i} \right)\),
where the latter inequality is satisfied because
\(G_{- A}\left( s_{A},{\tilde{S}}_{- A} \right)\) is a prisoner's
dilemma, and at \(s_{- i}\), they all choose the most cooperative strategy.
By inequality and by definition of
\(r_{i,k}^{*},u_{i}\left( s^{*} \right) \leq u_{i}^{*}\left( k_{i,1},s_{- i} \right) \Rightarrow u_{i}^{*}\left( k_{i},{\tilde{s}}_{- i} \right) \leq u_{i}^{*}\left( k_{i,1},s_{- i} \right)\),
where if \(1 \leq i \leq \tilde{m}\),
the inequality is strict, and therefore the result is still valid.
Finally, if the deviating coalition consists of \(m - 1\) players, for
each player \(i\) who deviates,
\({\tilde{s}}_{- i} \neq s_{- i}^{*}\), so
\(u_{i}^{*}\left( k,{\tilde{s}}_{- i} \right) \leq u_{i}^{*}\left( k_{i,1},s_{- i} \right)\)
as we have seen, and as before, the inequality is strict if
\(1 \leq i \leq \tilde{m}\). For the single player \(i\) who does
not deviate, we have that
\(u_{i}^{*}\left( k_{i,1},{\tilde{s}}_{- i} \right) \leq u_{i}^{*}\left( k_{i,1},s_{- i} \right)\)
as we also saw before: in this case we do not need the inequality
\(u_{i}^{*}\left( k,s_{- i} \right) \leq u_{i}^{*}\left( k - 1,s_{- i} \right)\)
if \(s_{- i} \neq s_{- i}^{*}\) (it would not work because
\(s_{- i} = s_{- i}^{*}\)) that we have been using, since \(i\) does
not deviate; therefore, it chooses the same strategy \(s_{i,k_{i,1}}\)
at \(s = \left( s_{1,k_{1,1}},\ldots s_{m,k_{m,1}},s_{- A} \right)\)
and at \(\tilde{s}\). Moreover, as always, inequality is strict if
\(1 \leq i \leq \tilde{m}\). We already analyzed all possible
coalitions for all possible \(s_{A} \in S_{A}\) with arbitrary
\(A \subseteq \{ 1,\ldots,n\}\) and
\({\tilde{S}}_{- A} \subseteq S_{- A}\) (where
\(1 \leq |A| \leq n - 2\)). Thus, we conclude that joint cooperation
is a strong Nash equilibrium in every restricted game of \(G\) in
which at least one \(r_{i,k}^{*} = r_{i,k}\) exists when all other players choose their
most cooperative strategy within the game. Specifically, given
\(k > 2\), \(u_{i}(k,1,\ldots,1) - r_{i,k}^{*} \leq u_{i}(2,1,\ldots,1) - r_{i,2}^{*} < u_{i}(E)\)
for all possible values of \(r_{i,k}^{*},r_{i,2}^{*}\) (with the
latter strict inequality since if \(r_{i,2}^{*} \neq r_{i,2}\),
likewise \(u_{i}\left( s^{*} \right) < u_{i}(E)\) with strict
inequality), so with a deviation from \(E\) of any number of players,
\(u_{i}^{*}\left( k,{\tilde{s}}_{- i} \right) < u_{i}(E) = u_{i}^{*}(E)\) (since
\(r_{i,1}^{*} = r_{i,1} = 0 \)) \(\forall 1 \leq i \leq n,1 \leq k \leq k_{i}\)
(with \(k = 1\) is satisfied by Lemma 3), so given any subset
\(B \subseteq \{ 1,\ldots,n\}\) of players,
\(\sum_{i \in B}^{}\mspace{2mu} u_{i}^{*}\left( P,E_{- B} \right) < \sum_{i \in B}^{}\mspace{2mu} u_{i}^{*}(E)\) \(\forall P \in S_{B} \setminus \left\{ E_{B} \right\}\).
This means that \(E\) is a strong Nash equilibrium in \(G\). $\square$

\vspace{1\baselineskip}

\textbf{Remark 4}: There is a case that we are not taking into account, because each restricted game has at least two players, but it could happen that there is a restriction on all strategies except the least cooperative one, for all players except for one. But in this case, if the player in question chooses to cooperate with any level other than the last one, due to the compensation, we must take by default his least cooperative strategy, so each player $i$ receives the same $u_i(s^*)$ as in the equilibrium of the game $G$. By the inequalities obtained in the proof of Theorem 2, we have that for the restricted games $\tilde{G}$ in which $u_i^*(s) \leq u_i(s^*)$ $\forall s \in S$ and for each player $i$ of $\tilde{G}$, joint cooperation is a strong (non-strict) Nash equilibrium, in which each player $i$ of $\tilde{G}$ obtains a utility of $u_i(s^*)$. Then, by cooperating, no player obtains lower utility than in the equilibrium of $G$, and therefore signing $C$ is a dominant strategy.

\vspace{1\baselineskip}

Assuming that a unit of utility is the minimum significant amount for which a player would change their strategy, we can define each \(\varepsilon_{i,k}\) in Theorems 1 and 2 as \(\varepsilon_{i,k} = k - 1\). These values satisfy the conditions \(\left( \varepsilon_{i,k} > \varepsilon_{i,k - 1} > \ldots > \varepsilon_{i,2} > 0 \right)\) and encourage players to cooperate, as they are significant and cooperation constitutes a strong equilibrium. At the same time, they minimize the utility loss if their constraints do not allow for cooperation. In this way, the values of \(r_{i,k}\) are defined as \(r_{i,k} := \sum_{m = 2}^{k} \left[ \max_{{\tilde{s}}_{- i} \in S_{- i}} \{ u_{i}\left( m,{\tilde{s}}_{- i} \right) - u_{i}\left( m - 1,{\tilde{s}}_{- i} \right) \} \right] + k - 1\), and our losing contracts are thus fully established by definition.

\subsection{Weak version}

Let us remember the public goods game: we have $n$ players where each can contribute a certain amount (by choosing a strategy among those available) to a common fund financing the public good, from which everyone benefits. This game is similar to the prisoner's dilemma because players obtain a higher utility if the public good is achieved (even after paying their contribution) than if they do not cooperate (defect) and the good is not produced.

The main difference with our prisoner's dilemma is that in the public goods game, a minimum collective cooperation is required for the good to be produced. If this minimum value is not reached, the players are not in a prisoner's dilemma because, regardless of their individual cooperation, the fund lacks the necessary financing to provide the good. This minimum value of collective cooperation is commonly referred to as the game's threshold. In our model, we assume that if this threshold is not met, players recover all of their contributions. However, if the threshold is exceeded, the surplus is utilized to improve the quality of the good, thereby providing greater benefits.

We are interested in representing the public goods game. To that end, the following definitions will be provided, which will establish a sequence of minimum thresholds depending on the level of collective cooperation.

\vspace{1\baselineskip}

\noindent \textbf{Definition 9.} Let \(G\) be an $n$-player game in which each player \(i\) has
\(k_{i}\) strategies. For each \(1 \leq i \leq n,1 \leq k \leq k_{i}\),
we define \(c_{i,k}\) as the contribution that the player \(i\) makes to
the common fund at \(G\) by choosing strategy \(k\), such that
\(0 = c_{i,k_{i}} < c_{i,k_{i} - 1} < \ldots < c_{i,1}\) \(\forall 1 \leq i \leq n\).

\vspace{1\baselineskip}

\noindent \textbf{Definition 10.} Let \(G\) be an $n$-player game in which each player \(i\) has \(k_{i}\) strategies. We say that \(G\) is a public goods game of order \(c\) (denoted by \(G = G(c)\)) if every restricted game \(G_{- A}\left( s_{A},{\tilde{S}}_{- A} \right)\) is a prisoner's dilemma, provided that the sum of the contributions from the strategies in \(s_{A}\) and the least cooperative strategies in \({\tilde{S}}_{- A}\) for the players in \(- A\) exceeds the minimum threshold \(c > 0\). We also will require that the utility of each player \(i\) in \(G(c)\) be equal to \(u_{i}\left( s^{*} \right)\) whenever \(\sum_{i = 1}^{n} c_{i,k} < c\)

\vspace{1\baselineskip}

Since players are refunded their contributions whenever the aggregate sum falls short of the minimum threshold, we can prove the following result, which is an extension of Theorem 2.

\vspace{1\baselineskip}

\noindent \textbf{Theorem 3.} \textit{Let \(G(c)\) be an $n$-player public goods game where each player
\(i\) has \(k_{i}\) strategies, and let \(C\) be a losing contract such that
$r_{i,k}^{*} = \sum_{m = 2}^{k}\mspace{2mu}\max_{{\tilde{s}}_{- i} \in S_{- i}}$ $\{ u_{i}\left( m,{\tilde{s}}_{- i} \right)-u_{i}\left( m - 1,{\tilde{s}}_{- i} \right)\} + \varepsilon_{i,k}$ (with $\varepsilon_{i,k} > \varepsilon_{i,k - 1} > \ldots > \varepsilon_{i,2} > 0)$
for each \(1 \leq i \leq n,2 \leq k \leq k_{i}\). Then, each
\(r_{i,k}^{*}\) is in its feasible region and \(C\) optimizes
\(E = \left( s_{1,1},\ldots,s_{n,1} \right)\), which is a strong Nash equilibrium. Moreover, \(C\) ensures
that in every restricted game containing at least one
\(r_{i,k}^{*} = r_{i,k}\) when all other players choose to cooperate, joint cooperation is a strong Nash equilibrium. In
addition, signing \(C\) is a dominant strategy for all players.}

\noindent \textit{Proof.} The proof that each \(r_{i,k}^{*}\) is in its feasible region and
that \(C\) optimizes \(E\) are similar to those of Theorem 2. Let
\(\tilde{G}(c)\) be a restricted game of \(G(c)\) in which
\(\exists j\) player of \(\tilde{G}(c)\) such that if \(s_{j,k}\) is
the most cooperative strategy in the latter, then
\(r_{j,k}^{*} = r_{j,k}\) if the other players of \(\tilde{G}(c)\)
also choose their most cooperative strategies in this last game. Without
loss of generality, suppose that each player \(\tilde{i}\) outside
\(\tilde{G}(c)\) chooses \(s_{\tilde{i},k}\), and also that the
least cooperative strategy of each player \(\tilde{i}\) in
\(\tilde{G}(c)\) is \(s_{\tilde{i},k}\). We have that if
\(\sum_{i = 1}^{n}\mspace{2mu} c_{i,k} \geq c\), \(\tilde{G}(c)\) is a
prisoner's dilemma, and by a test analogous to that of Theorem 2, joint
cooperation is a strong Nash equilibrium in this restricted game after
signing \(C\). Moreover, if
\(\sum_{i = 1}^{n}\mspace{2mu} c_{i,k} < c\), we have two options: there
is a restricted game of \(\tilde{G}(c)\) which is a prisoner's
dilemma or there is none at all. If we are in the first option, joint cooperation is a strong equilibrium due to a reasoning analogous to the previous one.
Otherwise, given any \(\tilde{j} \neq j\) player of
\(\tilde{G}(c)\), we consider the restricted game \(G^{*}(c)\) of
\(\tilde{G}(c)\) between \(\tilde{j}\) and \(j\) in which the
other players of \(\tilde{G}(c)\) choose their most cooperative
strategy within \(\tilde{G}(c)\), and in which \(j\) and
\(\tilde{j}\) dispose of their two most cooperative strategies among
those available in \(\tilde{G}(c)\). Since \(G^{*}(c)\) is not a
prisoner's dilemma, if we call again \(s_{i^{*},k}\) to the strategies
chosen by each player \(i^{*}\) outside \(G^{*}(c)\) (taking into
account that the strategies \(s_{i^{*},k}\) were the most cooperative in
\(\tilde{G}(c)\) for the players \(i^{*}\) who were in
\(\tilde{G}(c))\), and if we call
\(s_{l,\tilde{1}},s_{l,\tilde{2}}\) (for \(l = j,\tilde{j})\)
to the most cooperative and second most cooperative strategies of \(l\)
inside \(\tilde{G}(c)\) respectively, by Definition 10, we have
that
\(\sum_{i^{*} \neq j,\tilde{j}}^{}\mspace{2mu} c_{i^{*},k} + c_{j,\tilde{2}} + c_{\tilde{j},\tilde{2}} < c\),
so if \(j\) chooses \(s_{j,\tilde{2}}\) and \(\tilde{j}\)
chooses \(s_{\tilde{j},\tilde{2}}\), the utility will be equal
to \(u_{i}\left( s^{*} \right)\) for each \(1 \leq i \leq n\). Moreover,
if \(j\) and \(\tilde{j}\) choose \(s_{j,\tilde{1}}\) and
\(s_{\tilde{j},\tilde{2}}\) respectively, or if they choose
\(s_{j,\tilde{2}}\) and \(s_{\tilde{j},\tilde{1}}\)
respectively, we have two options: the total sum of contributions still
does not reach the minimum threshold \(c\), and thus each player
\(i\ (\)with \(1 \leq i \leq n)\) gets a utility of
\(u_{i}\left( s^{*} \right)\) as seen before, or now the total sum
reaches \(c\). However, in the latter case, owing to a test analogous to
that of Theorem 2,
\(u_{i}^{*}\left( \tilde{k},{\tilde{s}}_{- i} \right) \leq u_{i}^{*}\left( k,s_{- i} \right)\)
after signing \(C\) if a subset of players choose less or equally
cooperative strategies in \({\tilde{s}}_{- i}\) than in \(s_{- i}\)
and \(\tilde{k} \geq k\), where this inequality is strict for player
\(j\); because if all players choose their most cooperative strategies of those
available in \(\tilde{G}(c)\), \(r_{j,k}^{*} = r_{j,k}\) as stated. This
means that if one player cooperates and another does not in \(G^{*}(c)\), after signing \(C\), the sum of utilities of the players in \(\tilde{G}(c)\) is less than if both cooperated. In either case, it is proved that for any subset
of players from \(\tilde{G}(c)\) deviating from joint cooperation,
the total sum of their utilities after signing \(C\) will be lower
than if they all choose their most cooperative strategy among
\(\tilde{G}(c)\); thus, joint cooperation is a strong Nash
equilibrium under transferable utility. Specifically, given that
\(\sum_{i = 1}^{n}\mspace{2mu} c_{i,1} \geq c\) (since otherwise the
game would be meaningless), again by a proof analogous to that of
Theorem 2, \(E\) is a strong Nash equilibrium of \(G\). Moreover,
considering Remark 4, joint cooperation in all other restricted
games is a strong (non-strict) Nash equilibrium, in which each player
obtains an utility equal to the one that would be obtained in the
equilibrium of the original game. Therefore, there are incentives to
cooperate, and signing \(C\) is a dominant strategy. $\square$

\vspace{1\baselineskip}

In the literature, an $n$-player public goods game is represented by the
player's utility \(i\) given by
\(u_{i}\left( k,s_{- i} \right) = c_{i,1} - c_{i,k} + \frac{a}{n}\sum_{j = 1}^{n}\mspace{2mu} c_{j,k}\),
where \(1 < a < n\). In our definition, if we consider that the game is
symmetric in the sense that
\(c_{i,k} - c_{i,k + 1} = c_{j,\tilde{k}} - c_{j,\tilde{k} + 1}\forall 1 \leq i,j \leq n,1 \leq k \leq k_{i},1 \leq \tilde{k} \leq k_{j}\)
and that the utility of each player does not depend on who made the
contributions in the total sum; since any restricted game of two
players with two contiguous strategies each is a prisoner's dilemma (if
the minimum threshold is reached), we are implicitly assuming that
\(1 < \frac{2a}{n} \Leftrightarrow \frac{n}{2} < a\), because the player
values two units of contribution in the common fund more than one unit
kept for himself. If contributions are not so highly valued in the game
(\(a \leq \frac{n}{2}\)), besides the fact that our equilibrium analysis might not work for small coalitions, it wouldn't make sense for it to work either: before signing the contract,
there are restricted games in which players prefer to defect all rather
than cooperate together. Thus, in the face of certain restrictions,
there would be no reason to create incentives to cooperate: it is not
even convenient in the original game.

Continuing with the hypothesis of the game symmetry, further assuming
that the variation of utility is constant with respect to the strategies
of the others, we have that the amounts of Theorem 3 become
$r_{i,k} = \sum_{m = 2}^{k}\mspace{2mu}\lbrack \max_{{\tilde{s}}_{- i} \in S_{- i}}\{ u_{i}( m,$ ${\tilde{s}}_{- i}) - u_{i}\left( m - 1,{\tilde{s}}_{- i} \right)\} + \varepsilon_{i,k} = \sum_{m = 2}^{k}\mspace{2mu}\left\lbrack u_{i}\left( m,s_{- i} \right) - u_{i}\left( m - 1,s_{- i} \right) \right\rbrack + \varepsilon_{i,k} = u_{i}\left( k,s_{- i} \right) - u_{i}\left( 1,s_{- i} \right) + \varepsilon_{i,k}$,
which leaves us with
\(u_{i}^{*}\left( k,s_{- i} \right) = u_{i}\left( 1,s_{- i} \right) - \varepsilon_{i,k}\) if $r_{i,k}^*=r_{i,k}$,
where $s_{-i}$ could be any joint strategy of the rest of the players, since we
said that the variation of utility does not depend on the latter.
Furthermore, if this were not the case, the amounts defined as \(\tilde{r}_{i,k} = u_{i}^{*}\left( k,s_{- i} \right) = u_{i}\left( 1,s_{- i} \right) - \varepsilon_{i,k}\) would satisfy the three theorems stated above and would also yield higher payoffs than the $r_{i,k}$ previously used in the theorems (assuming the same $\varepsilon_{i,k}$ in both cases), in the sense
that players sacrifice less utility with $\tilde{r}_{i,k}$ than with $r_{i,k}$. The
difference is that the $\tilde{r}_{i,k}$ does depend on the joint strategy chosen by the
others when the game is not symmetrical. So, the context of the game
will determine which of the two amounts to use and whether or not
amounts depending on other strategies are allowed in the contract.

We proved that signing the contract is a dominant strategy, but we did
so assuming that the others also sign it: in reality, players can
speculate that the others will sign it and choose not to sign so as not
to contribute without incurring any loss. One way to resolve this is to make the contract effective only if all players sign it. Consequently, speculation is no longer possible, and signing the contract becomes a dominant strategy regardless of what others do. On the other hand, if we do not do so, not signing is not a dominant strategy either; therefore, executing it remains a valid option. We must clarify in the contract which of
the two options to choose. There is another problem: in the public goods
game, the utility received may not be transferable. In cases in which
the asset is an investment, we do not have this problem, but if it is an
asset that does not provide economic benefits, there may be players
without sufficient resources available to make the payoffs agreed upon
in the contract. In this case, these payoffs can be treated as debt, provided that all players agree to these terms before signing.

\section{Conclusions}

Losing contracts should be less effective than exchange
contracts. The latter are characterized by the absence of resource
wastage, a fundamental attribute that is absent in the former. However,
our analysis demonstrates that this is not the case. We found our
contracts to be more effective in every way; by paying the price of
wasting resources in the game, we were able to not only optimize the
cooperative equilibrium but also ensure that the latter is unique and
strong in any $n$-player prisoner's dilemma.

It was also determined that this games with $n$ players exhibit the two
properties of classical two-player dilemmas: a preference for the
cooperation of others and a primary conflict between individual interest
and collective benefits. This is due to the fact that as in two-player
dilemmas, joint cooperation is still more advantageous than defection,
despite the prevalence of defection. This supports the idea that our contracts incentivize cooperation both in the game and in its restricted games (since they are also dilemmas), allowing us to extend our results to games with potential constraints.

The structure that the $n$-player prisoner's dilemma establishes by
incorpora\-ting a minimum threshold is analogous to a minimum threshold in
a public goods game with sufficient valuation toward contributions to
the common fund. This enables us to further extend our results on this
type of games, which are used to study economic behavior. Beyond
applications, this structure also defines a theoretical framework for a
more rigorous analysis of the game, allowing us to conduct a local study of the strategic scenarios.

Returning to losing contracts, they are established by definition. Thus, unlike exchange contracts, the preplay stage reduces to a binary decision. If the contract becomes effective only when all players sign it, then signing it is a dominant strategy for everyone. In this way, our contracts achieve a resolution of the dilemma: the latter is solved by the losing contract defined in our theorems, as it induces a situation where all players obtain a higher payoff than in the equilibrium of the original game. This result also holds in the presence of potential constraints.

This work could be extended in future research in three different directions. The first consists of finding preplay contracts that optimize the equilibrium in each restricted game of the public goods games with lower valuation of contributions to the common fund. The second is the analysis of other games through their internal structures and their restricted games, from which new information could be obtained. The third is the study of our losing contracts in other types of games, since there may also be other games that can be solved by losing contracts before play even starts.

\vspace{1\baselineskip}

\section*{Acknowledgements}

I would like to thank the Department of Mathematics at the University of Buenos Aires for the academic resources and support provided during my studies. I am especially grateful to Professor Ariel Arbiser for introducing me to the field of Game Theory. I also wish to express my gratitude to Eric Brandwein for providing the endorsement required to publish this paper. Finally, I thank Enago for their professional English translation services.

\bibliographystyle{plainnat} 
\bibliography{referencias}

\end{document}